\documentclass[aps,prd,preprint,tightenlines,superscriptaddress,showpacs,byrevtex]{revtex4}  

\usepackage{graphicx} 
\usepackage{dcolumn}  
\usepackage{bm}
\usepackage{amsmath}
\usepackage{color}
\definecolor{darkgreen}{rgb}{0.,0.55,0.}
\definecolor{orange}{rgb}{0.75,0.6,0.}
\graphicspath{{ps}}

\usepackage{lineno}

\newcommand{\Brr}{\ensuremath{B^{0} \to \rho^{0}\rho^{0}}}
\newcommand{\Bfr}{\ensuremath{B^{0} \to f_{0}\rho^{0}}}
\newcommand{\Bff}{\ensuremath{B^{0} \to f_{0}f_{0}}}
\newcommand{\Bfpp}{\ensuremath{B^{0} \to f_{0}\pi^{+}\pi^{-}}}
\newcommand{\Brpp}{\ensuremath{B^{0} \to \rho^{0}\pi^{+}\pi^{-}}}
\newcommand{\Bpppp}{\ensuremath{B^{0} \to \pi^{+}\pi^{-}\pi^{+}\pi^{-}}}
\newcommand{\Fpp}{\ensuremath{f_{0} \to \pi^{+}\pi^{-}}}
\newcommand{\aonepi}{\ensuremath{B^{0} \to a^{\pm}_{1}\pi^{\mp}}}
\newcommand{\bonepi}{\ensuremath{B^{0} \to b^{\pm}_{1}\pi^{\mp}}}
\newcommand{\atwopi}{\ensuremath{B^{0} \to a^{\pm}_{2}\pi^{\mp}}}

\newcommand{\dpi}{\ensuremath{B^{0} \to D^{-} [K^{+}\pi^{-}\pi^{-}] \pi^{+}}}

\newcommand{\epem}{\ensuremath{e^{+} e^{-}}}
\newcommand{\qqbar}{\ensuremath{q \bar{q}}}
\newcommand{\BBbar}{\ensuremath{B \bar{B}}}
\newcommand{\BzBzb}{\ensuremath{B^{0} \bar{B}^{0}}}
\newcommand{\BpBm}{\ensuremath{B^{+} B^{-}}}

\newcommand{\pip}{\ensuremath{\pi^{+}}}
\newcommand{\pim}{\ensuremath{\pi^{-}}}

\newcommand{\rz}{\ensuremath{\rho^{0}}}

\newcommand{\aone}{\ensuremath{a_{1}^{\pm}}}
\newcommand{\atwo}{\ensuremath{a_{2}^{\pm}}}
\newcommand{\bone}{\ensuremath{b_{1}^{\pm}}}
\newcommand{\Ks}{\ensuremath{K^{0}_{S}}}
\newcommand{\Bz}{\ensuremath{B^{0}}}

\newcommand{\Ups}{\ensuremath{\Upsilon(4S)}}

\newcommand{\Btag}{\ensuremath{B^{0}_{\rm Tag}}}

\newcommand{\Mbc}{\ensuremath{M_{\rm bc}}}
\newcommand{\De}{\ensuremath{\Delta E}}
\newcommand{\Fsb}{\ensuremath{{\cal F}_{S/B}}}

\newcommand{\Mpp}{\ensuremath{m_{\pi^{+}\pi^{-}}}}
\newcommand{\cH}{\ensuremath{\cos\theta_{H}}}

\newcommand{\phitwo}{\ensuremath{\phi_{2}}}
\newcommand{\phitwoeff}{\ensuremath{\phi^{\rm eff}_{2}}}

\begin{document}


\title{ \quad\\[0.5cm]  Study of $\bm{\Brr}$ decays, implications for the CKM angle $\bm{\phi_2}$ and search for other $\bm{B^0}$ decay modes with a four-pion final state}
\noaffiliation
\affiliation{University of the Basque Country UPV/EHU, 48080 Bilbao}
\affiliation{Beihang University, Beijing 100191}
\affiliation{Budker Institute of Nuclear Physics SB RAS and Novosibirsk State University, Novosibirsk 630090}
\affiliation{Faculty of Mathematics and Physics, Charles University, 121 16 Prague}
\affiliation{Chiba University, Chiba 263-8522}
\affiliation{University of Cincinnati, Cincinnati, Ohio 45221}
\affiliation{Deutsches Elektronen--Synchrotron, 22607 Hamburg}
\affiliation{Department of Physics, Fu Jen Catholic University, Taipei 24205}
\affiliation{Justus-Liebig-Universit\"at Gie\ss{}en, 35392 Gie\ss{}en}
\affiliation{Hanyang University, Seoul 133-791}
\affiliation{University of Hawaii, Honolulu, Hawaii 96822}
\affiliation{High Energy Accelerator Research Organization (KEK), Tsukuba 305-0801}
\affiliation{Ikerbasque, 48011 Bilbao}
\affiliation{Indian Institute of Technology Guwahati, Assam 781039}
\affiliation{Indian Institute of Technology Madras, Chennai 600036}
\affiliation{Institute of High Energy Physics, Chinese Academy of Sciences, Beijing 100049}
\affiliation{Institute of High Energy Physics, Vienna 1050}
\affiliation{Institute for High Energy Physics, Protvino 142281}
\affiliation{INFN - Sezione di Torino, 10125 Torino}
\affiliation{Institute for Theoretical and Experimental Physics, Moscow 117218}
\affiliation{J. Stefan Institute, 1000 Ljubljana}
\affiliation{Kanagawa University, Yokohama 221-8686}
\affiliation{Institut f\"ur Experimentelle Kernphysik, Karlsruher Institut f\"ur Technologie, 76131 Karlsruhe}
\affiliation{Kavli Institute for the Physics and Mathematics of the Universe (WPI), University of Tokyo, Kashiwa 277-8583}
\affiliation{Korea Institute of Science and Technology Information, Daejeon 305-806}
\affiliation{Korea University, Seoul 136-713}
\affiliation{Kyungpook National University, Daegu 702-701}
\affiliation{\'Ecole Polytechnique F\'ed\'erale de Lausanne (EPFL), Lausanne 1015}
\affiliation{Faculty of Mathematics and Physics, University of Ljubljana, 1000 Ljubljana}
\affiliation{Luther College, Decorah, Iowa 52101}
\affiliation{University of Maribor, 2000 Maribor}
\affiliation{Max-Planck-Institut f\"ur Physik, 80805 M\"unchen}
\affiliation{School of Physics, University of Melbourne, Victoria 3010}
\affiliation{Moscow Physical Engineering Institute, Moscow 115409}
\affiliation{Graduate School of Science, Nagoya University, Nagoya 464-8602}
\affiliation{Kobayashi-Maskawa Institute, Nagoya University, Nagoya 464-8602}
\affiliation{Nara Women's University, Nara 630-8506}
\affiliation{National Central University, Chung-li 32054}
\affiliation{National United University, Miao Li 36003}
\affiliation{Department of Physics, National Taiwan University, Taipei 10617}
\affiliation{H. Niewodniczanski Institute of Nuclear Physics, Krakow 31-342}
\affiliation{Nippon Dental University, Niigata 951-8580}
\affiliation{Niigata University, Niigata 950-2181}
\affiliation{Osaka City University, Osaka 558-8585}
\affiliation{Pacific Northwest National Laboratory, Richland, Washington 99352}
\affiliation{Panjab University, Chandigarh 160014}
\affiliation{University of Pittsburgh, Pittsburgh, Pennsylvania 15260}
\affiliation{University of Science and Technology of China, Hefei 230026}
\affiliation{Seoul National University, Seoul 151-742}
\affiliation{Soongsil University, Seoul 156-743}
\affiliation{Sungkyunkwan University, Suwon 440-746}
\affiliation{School of Physics, University of Sydney, NSW 2006}
\affiliation{Tata Institute of Fundamental Research, Mumbai 400005}
\affiliation{Excellence Cluster Universe, Technische Universit\"at M\"unchen, 85748 Garching}
\affiliation{Toho University, Funabashi 274-8510}
\affiliation{Tohoku Gakuin University, Tagajo 985-8537}
\affiliation{Tohoku University, Sendai 980-8578}
\affiliation{Department of Physics, University of Tokyo, Tokyo 113-0033}
\affiliation{Tokyo Institute of Technology, Tokyo 152-8550}
\affiliation{Tokyo Metropolitan University, Tokyo 192-0397}
\affiliation{Tokyo University of Agriculture and Technology, Tokyo 184-8588}
\affiliation{CNP, Virginia Polytechnic Institute and State University, Blacksburg, Virginia 24061}
\affiliation{Wayne State University, Detroit, Michigan 48202}
\affiliation{Yamagata University, Yamagata 990-8560}
\affiliation{Yonsei University, Seoul 120-749}
 \author{P.~Vanhoefer}\affiliation{Max-Planck-Institut f\"ur Physik, 80805 M\"unchen} 
  \author{J.~Dalseno}\affiliation{Max-Planck-Institut f\"ur Physik, 80805 M\"unchen}\affiliation{Excellence Cluster Universe, Technische Universit\"at M\"unchen, 85748 Garching} 
  \author{C.~Kiesling}\affiliation{Max-Planck-Institut f\"ur Physik, 80805 M\"unchen} 

  \author{I.~Adachi}\affiliation{High Energy Accelerator Research Organization (KEK), Tsukuba 305-0801} 
  \author{H.~Aihara}\affiliation{Department of Physics, University of Tokyo, Tokyo 113-0033} 
  \author{D.~M.~Asner}\affiliation{Pacific Northwest National Laboratory, Richland, Washington 99352} 
  \author{V.~Aulchenko}\affiliation{Budker Institute of Nuclear Physics SB RAS and Novosibirsk State University, Novosibirsk 630090} 
  \author{T.~Aushev}\affiliation{Institute for Theoretical and Experimental Physics, Moscow 117218} 
  \author{A.~M.~Bakich}\affiliation{School of Physics, University of Sydney, NSW 2006} 
  \author{A.~Bala}\affiliation{Panjab University, Chandigarh 160014} 
% ...
\collaboration{The Belle Collaboration}

\begin{abstract}
   We present a study of the branching fraction of the decay \Brr\ and the fraction of longitudinally polarized \rz\ mesons in this decay. The results are obtained from the final data sample containing $772 \times 10^{6}$ \BBbar\ pairs collected at the \Ups\ resonance with the Belle detector at the KEKB asymmetric-energy \epem\ collider. We find $166 \pm 59$ \Brr\ events (including systematic uncertainties), corresponding to a branching fraction of 
\begin{center} 
${\cal B}(\Brr) = (1.02\pm 0.30\;(\rm stat)  \pm 0.15\;(\rm syst))\times 10^{-6} $ 
\end{center} 
with a significance of $3.4$ standard deviations and a longitudinal polarization fraction 
\begin{center}   
$f_L = 0.21^{+0.18}_{-0.22} \;(\rm stat) \pm 0.15 \;(\rm syst)$.
\end{center} 
We use the longitudinal polarization fraction to determine the Cabibbo-Kobayashi-Maskawa matrix angle $\phi_2 = (84.9 \pm 13.5)^{\circ}$ through an isospin analysis in the $B\to\rho\rho$ system.
We furthermore find  $125 \pm 41$ \Bfr\ events, corresponding to
\begin{center}
${\cal B}(\Bfr) \times {\cal B}(\Fpp) = (0.78 \pm 0.22 \; (\rm stat) \pm 0.11 \;(\rm syst))\times 10^{-6}$,\\
  \end{center}
with a significance of $3.1$ standard deviations. We find no other significant contribution with the same final state, and set upper
   limits at 90\% confidence level on the (product) branching fractions,
 \begin{center}
${\cal B}(\Bpppp) < 11.2 \times 10^{-6} $,\\
${\cal B}(\Brpp) < 12.0 \times 10^{-6}$, \\
${\cal B}(\Bfpp) \times {\cal B}(\Fpp) < 3.0 \times 10^{-6}$ and\\
${\cal B}(\Bff) \times {\cal B}(\Fpp)^{2} < 0.2 \times 10^{-6}$. \\
 \end{center}
 
\end{abstract}


\maketitle

\tighten

{\renewcommand{\thefootnote}{\fnsymbol{footnote}}}
\setcounter{footnote}{0}

\section{Introduction}
$CP$ violation in the standard model (SM) is due to an irreducible complex phase in the Cabibbo-Kobayashi-Maskawa (CKM) quark-mixing matrix~\cite{Cabibbo,KM}. Mixing-induced $CP$ violation in the $B$ sector has been clearly observed by the Belle~\cite{jpsiks_Belle, jpsiks_Belle2} and BaBar~\cite{jpsiks_BABAR, jpsiks_BABAR2} collaborations in the $b \rightarrow c \bar c s$ induced decay $\Bz \rightarrow J/\psi \Ks$, while many other modes provide additional information on $CP$ violating parameters~\cite{BelleResults, CKMandUT}.

Decays that proceed predominantly through the $b \rightarrow u \bar u d$ transition are sensitive to one of the angles of the unitarity triangle, $\phitwo\;({\rm or}\;\alpha) \equiv \arg[(-V_{td}V^{*}_{tb})/(V_{ud}V^{*}_{ub})]$; its current world average is $\phi_2 = (88.5^{+4.7}_{-4.4})^{\circ}$~\cite{ckmfitter}. The Belle, BaBar and LHCb collaborations have reported time-dependent $CP$ asymmetries in these modes that include decays such as $\Bz \rightarrow \pip \pim$~\cite{pipi_Belle,pipi_BABAR,pipi_LHCb}, $\rho^{\pm} \pi^{\mp}$~\cite{rhopi_Belle,rhopi_BABAR}, $\rho^{+} \rho^{-}$~\cite{rhorho_Belle,rhorho_BABAR}, $\rho^{0} \rho^{0}$~\cite{r0r0_Belle,r0r0_BABAR} and $a_{1}^{\pm}\pi^{\mp}$~\cite{jeremy_a1pi,a1pi_BABAR1,a1pi_BABAR2}. A feature common to these measurements is that possible loop contributions, in addition to the leading-order tree amplitude, can shift the measured angle to $\phitwoeff\ \equiv  \phitwo\ + \Delta\phi_{2}$. This inconvenience can be overcome with bounds on $\Delta \phitwo$ determined using either an isospin analysis~\cite{theory_isospin} or $SU(3)$ flavor symmetry~\cite{theory_a1pi}.

\begin{figure}[b]
  \centering
  \includegraphics[height=120pt,width=!]{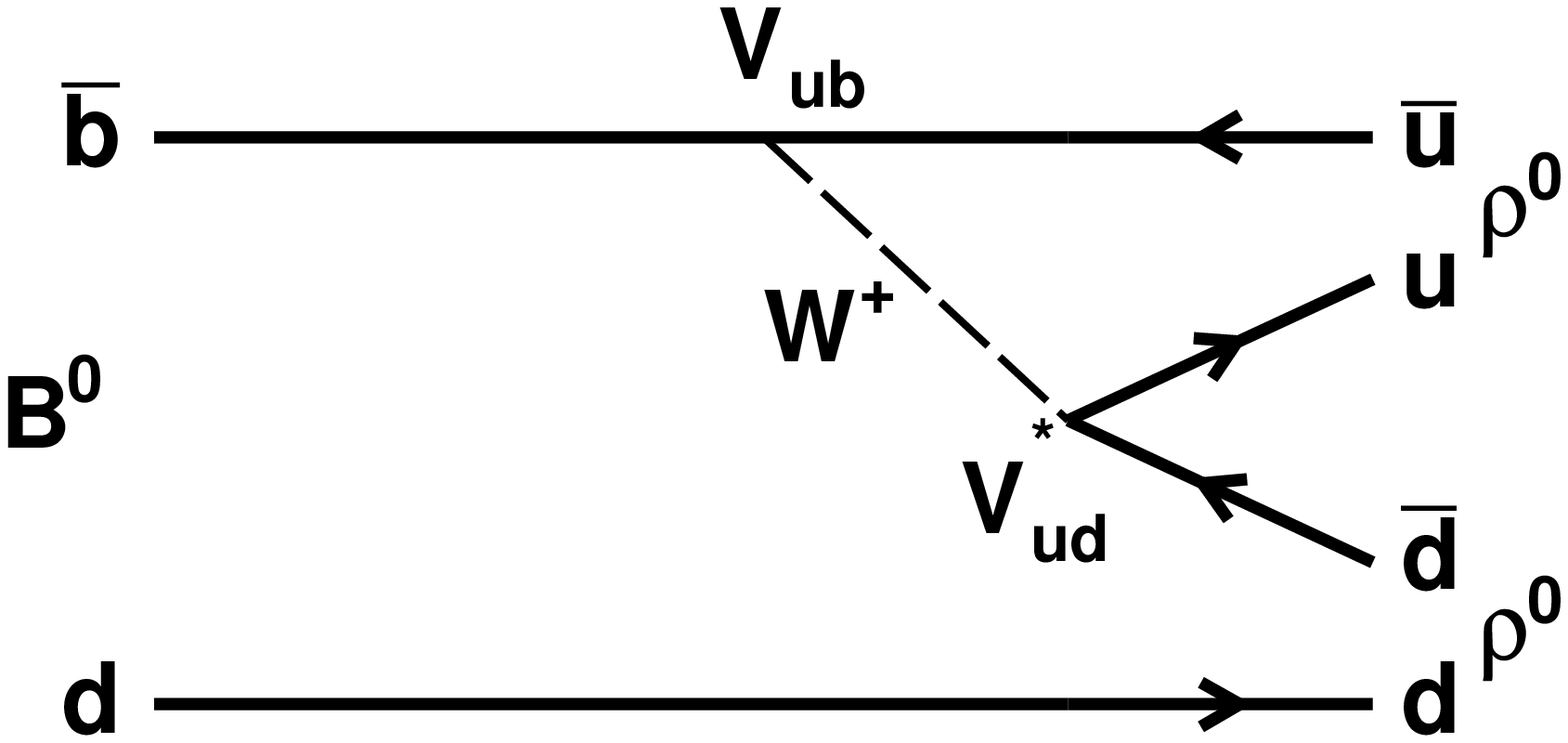}
  \includegraphics[height=120pt,width=!]{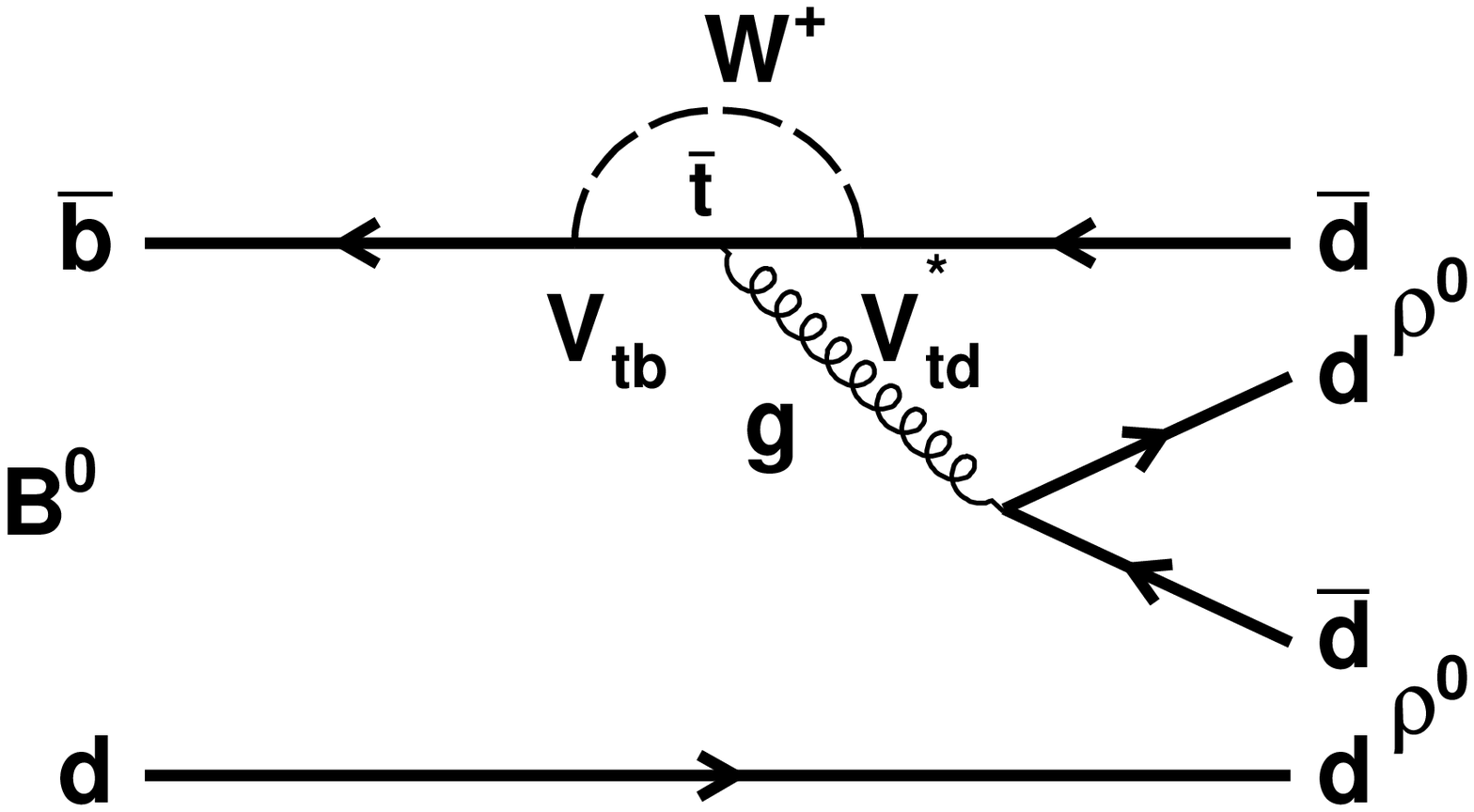}
  \caption{Leading-order tree (left) and penguin (right) diagrams for the decay \Brr.}
  \label{fig_r0r0}
\end{figure}

This analysis is concerned with the branching fraction of $B^0(\bar{B}^0)\to \rho^0\rho^0$ decays, the fraction of longitudinal polarization in these decays and further decays of the $B$ meson into four-charged-pion final states as the \rz\ decays into two charged pions. The leading-order tree and penguin diagrams of $\Brr$ decays are shown in Fig.~\ref{fig_r0r0}. 
Since the dominant tree process is color-suppressed, \Brr\ is expected to be less probable than its isospin partners. The SM, using perturbative QCD (pQCD) or QCD factorization in the heavy quark limit~\cite{pQCD, fQCD2, fQCD3, fQCD4, BVV, fQCD, BVV2}, predicts the \Brr\ branching fraction to be $\sim 1\times 10^{-6}$. 

\begin{figure}[b]
  \centering
  \includegraphics[height=120pt,width=!]{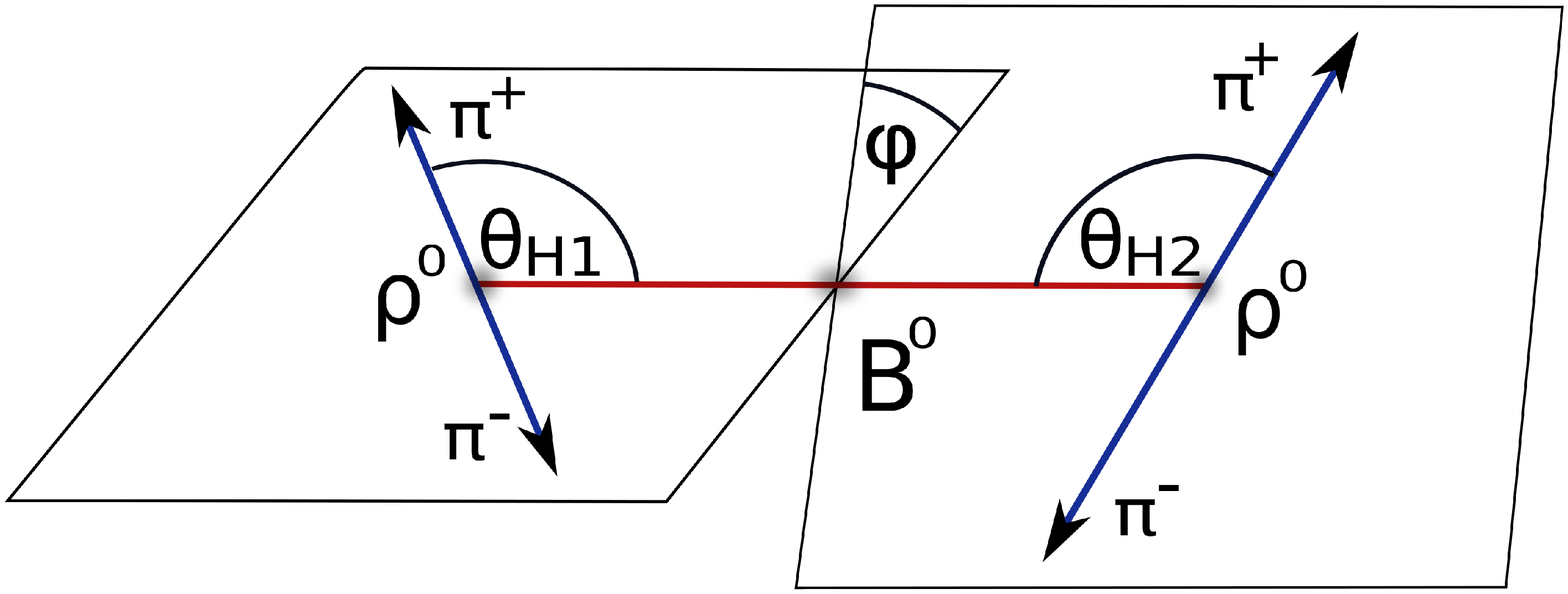}
  \caption{Definition of the helicity angles $\theta_{Hk}$ for each $\rho$, identified by index $k=1,2$. 
}
  \label{fig_helicity}
\end{figure}

The $\rz\rz$ vector-vector state is not a pure $CP$ eigenstate, but rather a superposition of $CP$-even and -odd states, or three helicity amplitudes, which can be separated through an angular analysis. We use the helicity basis where the angles $\theta_{Hk}|_{k=1,2}$, each defined as the angle between the $\pip$ and the $B$ flight directions in the rest frame of the $k^{\rm th}$ \rz\ (Fig.~\ref{fig_helicity}), can be used to separate longitudinally ($CP$-even) from transversely ($CP$-even and -odd) polarized \rz\ mesons.

Analogous to the decay of a $B^0$ meson into two charged $\rho$ mesons~\cite{rhorho_Belle,rhorho_BABAR}, \Brr\ is expected to decay predominantly into longitudinally polarized $\rz$s. However, color-suppressed $B$ meson decays into two vector particles are especially difficult to predict; one important difficulty, for example, is the non-factorization of the spectator-scattering for the transverse amplitude even at leading order~\cite{fQCD, BVV2}. Hence, this analysis provides an excellent test of the assumptions used in these frameworks and improves our understanding of the strong interaction.

In Sec.~\ref{Data Set And Belle Detector}, we briefly describe the data set and the Belle detector. The event selection and the model used for the branching fraction measurement are described in Secs.~\ref{Event Selection} and \ref{Event Model}, respectively, where in the latter section we also comment on differences with the previous Belle analysis. The fit result is presented in Sec.~\ref{Fit Result}, followed by validity checks in Sec.~\ref{Validity Checks and Significance}. The systematic uncertainties are discussed in Sec.~\ref{Systematic Uncertainties} and a constraint of the CKM phase \phitwo\ is presented in Sec.~\ref{phi2 constraint}, followed by a discussion of the result and our conclusion.

\section{Data Set And Belle Detector}
\label{Data Set And Belle Detector}
This measurement is based on the final data sample containing $772 \times 10^{6}$ \BBbar\ pairs collected with the Belle detector at the KEKB asymmetric-energy \epem\ ($3.5$ on $8~{\rm GeV}$) collider~\cite{KEKB}. At the \Ups\ resonance ($\sqrt{s}=10.58$~GeV), the Lorentz boost of the produced \BBbar\ pairs is $\beta\gamma=0.425$ along the $z$ direction, which is opposite the positron beam direction.

The Belle detector is a large-solid-angle magnetic
spectrometer that consists of a silicon vertex detector (SVD),
a 50-layer central drift chamber (CDC), an array of
aerogel threshold Cherenkov counters (ACC), 
a barrel-like arrangement of time-of-flight
scintillation counters (TOF), and an electromagnetic calorimeter
comprising of CsI(Tl) crystals (ECL) located inside 
a superconducting solenoid coil that provides a 1.5~T
magnetic field.  An iron flux-return located outside of
the coil is instrumented to detect $K_L^0$ mesons and to identify
muons (KLM).  The detector
is described in detail elsewhere~\cite{Belle}.
Two inner detector configurations were used. A 2.0 cm radius beampipe
and a 3-layer silicon vertex detector (SVD1) were used for the first sample
of $152 \times 10^6$ \BBbar\ pairs, while a 1.5 cm radius beampipe, a 4-layer
silicon detector (SVD2) and a small-cell inner drift chamber were used to record  
the remaining $620 \times 10^6$ \BBbar\ pairs~\cite{svd2}. We use a GEANT-based
 Monte Carlo (MC) simulation to model the response of the detector and determine
 its acceptance~\cite{GEANT}.

\section{Event Selection}
\label{Event Selection}
We reconstruct $\Brr$, where $\rz\to\pip\pim$. Charged tracks have to fulfill requirements on the distance of closest approach to the interaction point: $|dz| < 5.0 \; {\rm cm}$ and $dr < 0.5 \; {\rm cm}$ along and perpendicular to the $z$ axis, respectively. With information obtained from the CDC, ACC and TOF, particle identification (PID) is determined with the likelihood ratio ${\cal L}_{i/j}\equiv {\cal L}_{i}/({\cal L}_{i} + {\cal L}_{j})$, where ${\cal L}_{i}$ (${\cal L}_{j}$) is the likelihood that the particle is of type $i$ ($j$). We require ${\cal L}_{K/\pi} < 0.4$, which retains 90\% of all pions but only 10\% of all kaons. In addition, we place vetoes on particles consistent with the electron or proton hypotheses. Requirements of at least two hits in the z and one hit in the azimuthal strips of the SVD~\cite{ResFunc} are imposed on the charged tracks, to permit a subsequent measurement of the $CP$ asymmetries.

Intermediate dipion states are reconstructed above the \Ks\ region with an invariant mass $0.52 \; {\rm GeV}/c^{2} < m(\pip \pim) < 1.15 \; {\rm GeV}/c^{2}$ straddling the broad \rz(770) resonance~\cite{PDG}. This range retains $93\%$ of the phase-space available for a \rz\ coming from $\Brr$ decays. Upon combination of two dipion states, a $B$ candidate is formed. All remaining particles are associated with the accompanying $B$ meson in the event, referred to as \Btag.

Reconstructed $B$ candidates are described with two kinematic variables: the beam-energy-constrained mass $\Mbc \equiv \sqrt{(E^{\rm CMS}_{\rm beam}/c^2)^{2} - (p^{\rm CMS}_{B}/c)^{2}}$, and the energy difference $\De \equiv E^{\rm CMS}_{B} - E^{\rm CMS}_{\rm beam}$, where $E^{\rm CMS}_{\rm beam}$ is the beam energy and $E^{\rm CMS}_{B}$ ($p^{\rm CMS}_{B}$) is the energy (momentum) of the $B$ meson, evaluated in the center-of-mass system (CMS). The $B$ candidates that satisfy  $\Mbc > 5.27 \; {\rm GeV}/c^{2}$  and $|\De| < 0.1 \; {\rm GeV}$ are selected for further analysis.

\begin{table}
  \centering
  \caption{Summary of criteria to remove peaking background modes. A muon mass hypothesis has been applied to specific tracks for the $J/\psi$ channel. Here, $\pip_{1}\pim_{2}$ forms the first and $\pip_{3}\pim_{4}$ the second $\rz$ candidate of a reconstructed event. $X$ represents any intermediate state or track combination that leads to a 4-body final state. Only peaking combinations are vetoed.}
  \begin{tabular}
    {@{\hspace{0.5cm}}c@{\hspace{0.25cm}}  @{\hspace{0.25cm}}c@{\hspace{0.25cm}} }
    \hline \hline
    Regions vetoed & Modes vetoed \\
    \hline
 $1.85 \; {\rm GeV}/c^{2} < m(\pip\pim) < 1.89 \; {\rm GeV}/c^{2}$ & $B \rightarrow D^{0} [\pip\pim] X$ \\
    $1.85 \; {\rm GeV}/c^{2} < m(\pip_{1}\pim_{2}\pip_{3}) < 1.89 \; {\rm GeV}/c^{2}$ & $B \rightarrow D^{+} [\pip\pim\pip] X$ \\
  $1.85 \; {\rm GeV}/c^{2} < m(\pim_{2}\pip_{3}\pim_{4}) < 1.89 \; {\rm GeV}/c^{2}$ & $B \rightarrow D^{-} [\pim\pip\pim] X$ \\
  $1.95 \; {\rm GeV}/c^{2} < m(\pip_{1}\pim_{2}\pip_{3}) < 1.99 \; {\rm GeV}/c^{2}$ & $B \rightarrow D_{s}^{+} [\pip\pim\pip] X$ \\
 $1.95 \; {\rm GeV}/c^{2} < m(\pim_{2}\pip_{3}\pim_{4}) < 1.99 \; {\rm GeV}/c^{2}$ & $B \rightarrow D_{s}^{-} [\pim\pip\pim] X$ \\
    $3.06 \; {\rm GeV}/c^{2} < m(\mu^{+}\mu^{-}) < 3.14 \; {\rm GeV}/c^{2}$ & $B \rightarrow J/\psi [\mu^{+} \mu^{-}] X$ \\
    $0.478 \; {\rm GeV}/c^{2} < m(\pi^{-}_{2}\pi^{+}_{3}) < 0.512 \; {\rm GeV}/c^{2}$ & $B \rightarrow \Ks [\pip \pim] X$ \\
    \hline \hline
  \end{tabular}
  \label{tab_veto}
\end{table}

To reduce peaking background coming from charm ($b \to c$) decays of the $B$ meson with a similar final-state topology such as $B^{0} \to\ D^{-}[\pim\pip\pim]\pip$ or backgrounds due to particle mis-identification, we place vetoes on various combinations of the four charged tracks forming our $B^0$ candidate, as summarized in Table~\ref{tab_veto}. The total efficiency loss due to these vetoes is $4.4\%$ while the charm $B$ decays contributions are decreased by $20\%$ reducing their peaking contribution to a negligible level. 

\begin{figure}[h]
  \centering
  \includegraphics[height=300pt,width=!]{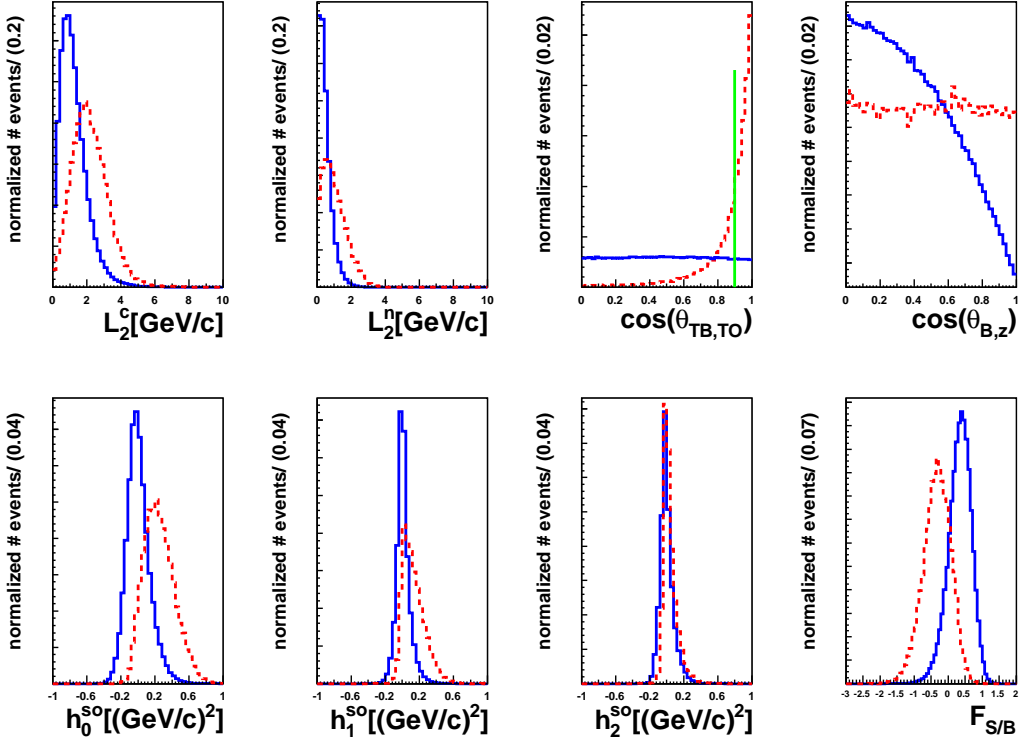}
  \caption{(color online) Simulated MC and off-resonance distributions for the quantities used
to construct the Fisher discriminant  \Fsb, which is shown in the lower right plot. The solid (blue) histograms show the distribution for \BBbar\ MC events, while the dashed (red) histograms show the distributions for events from off-resonance data, both normalized to the same area. The vertical line indicates the requirement of $|\cos \theta_{\rm TB, TO}| < 0.9$ being applied before the training of the fisher.}
  \label{fig_fd}
\end{figure}

The dominant background contribution comes from continuum ($\epem \rightarrow \qqbar$, where $q=u,d,s,c$) events. We use their jet-like topology to separate them from the more spherical \BBbar\ decays using a Fisher discriminant~\cite{fisher} \Fsb\ constructed from the following seven variables:
\begin{itemize}
\item $L^{c,n}_{2} \equiv \sum_{c,n} p^{\rm CMS}_{c,n}\cos^{2} \theta_{p^{\rm CMS}_{c,n},{\rm TB}}$, where the sum of the CMS momenta runs over the charged tracks ($c$) and neutral clusters ($n$) on the tag side; the angle is between the particle direction and $B$ thrust direction which points into the reconstructed $B$ meson's flight direction.
\item $\cos \theta_{\rm TB, TO}$, where the angle is between the $B$ thrust direction and the thrust of the tag side.
\item $\cos \theta_{B, z}$, where the angle is between the $B$ flight direction and the $z$ direction.
\item $h^{so}_{k} \equiv \sum_{i,j_{k}}|\vec{p}_i||\vec{p_{j_{k}}}| P_2(\cos(\theta_{ij_{k}}))|$, closely related to the Fox-Wolfram moments~\cite{foxw1, foxw2}, where $\vec{p_{i}}$ is the CMS momentum of the $i^{\rm th}$ track from the signal side ($s$), $\vec{p_{j_{k}}}$ is the CMS momentum of the $j_k^{\rm th}$ particle from the other side ($o$), $\theta_{ij_{k}}$ is the angle between particle $i$ and $j_k$ and $P_2$ is the second order Legendre polynomial. For the other side, we distinguish three cases using index $k=0,1$ and $2$, for charged tracks, neutral particles and missing energy (treated as a particle), respectively. 
\end{itemize}
The respective distributions and \Fsb\ are shown in Fig.~\ref{fig_fd}. 
We require $|\cos \theta_{\rm TB, TO}| < 0.9$ in order to reject 50\% of the continuum background while retaining 90\% of signal. The \BBbar\ and \qqbar\ training samples are taken from signal MC events and from (off-resonance) data taken below the \Ups\ resonance, respectively, and also fulfill $|\cos \theta_{\rm TB, TO}| < 0.9$. The Fisher discriminant is also required to satisfy $-3 < \Fsb < 2$. This, together with the previously mentioned requirements of $|\De|<0.1 \;{\rm GeV}$, $\Mbc > 5.27 \; {\rm GeV}/c^{2}$ and  $0.52 \; {\rm GeV}/c^{2} < m(\pip \pim) < 1.15 \; {\rm GeV}/c^{2}$, defines the fit region.

According to MC simulation, 1.06 $B$ candidates are reconstructed on average per signal event. Selecting the best $B$ candidate having \Mbc\ nearest the nominal $B$ meson mass~\cite{PDG}, the correct $B$ is chosen in $75\%$ of all events with multiple candidates. If both possible dipion combinations of the four pions fall within the fit region, the combination with the largest momentum difference between its pions is chosen. Since our selection criteria biases the \Mbc\ distributions, we do not use this variable in the fit to data. The fraction of mis-reconstructed signal events, where at least one pion is taken from the other $B$ meson, is found to be $9\%$.

We also perform a vertex fit~\cite{ResFunc} and employ the flavor tagging routine described in Ref.~\cite{Tagging}. The tagging information is represented by two parameters, the \Btag\ flavor $q=\pm1$ for $\Btag = B,\bar{B}$, and the tagging quality $r$. The parameter $r$ is an event-by-event, MC determined flavor-tagging dilution factor that ranges from $r = 0$ for no flavor discrimination to $r = 1$ for unambiguous flavor assignment. We find \Fsb\ to be correlated with $r$ and therefore divide the data into 7 $r$-bins, labeled with the index $l$.

We complete the reconstruction with a randomization of the ordering of the two $\rho^0$s by interchanging the two \rz\ candidates for every other event to avoid an artificial asymmetry in the distribution of the helicity angles arising mainly from momentum ordering in the reconstruction.

We study backgrounds coming from $b \rightarrow u \bar{u} d$ channels with the same final state as signal, namely $B^{0} \to  a_1^{\pm}\pi^{\mp}, \;a_2^{\pm}\pi^{\mp},\;b_1^{\pm}\pi^{\mp},\; f_0(980)\rz,\; $ $f_0(980)f_0(980),\; \rz\pip\pim,\; f_0(980)\pip\pim$ and $\pip\pim\pip\pim$. The detection efficiencies $\epsilon$ of the considered four-charged-pion final states after applying all mentioned selection criteria are listed in Table~\ref{t_eps}. Using an independent control sample, we determine a correction factor to the efficiency that account for the differences in particle identification between data and MC, $\eta=0.85\pm0.03$ for all four-pion modes.
\begin{table}
  \centering
  \caption{The reconstruction efficiencies ($\epsilon$) of the signal for the four-pion final states calculated from Monte Carlo. Here, the correction for the differences in particle identification between data and MC has not been applied. The errors are statistical. For \Brr, we use the $f_L$ value obtained from the fit to data.}
  \begin{tabular}
 {@{\hspace{0.5cm}}c@{\hspace{0.25cm}}  @{\hspace{0.25cm}}c@{\hspace{0.25cm}} @{\hspace{0.25cm}}c@{\hspace{0.25cm}}}
  \hline \hline
 Mode &  $\epsilon_{\rm SVD1}$ ($\%$) &  $\epsilon_{\rm SVD2}$ ($\%$) \\
\hline
\Brr & $22.38 \pm 0.05 $& $ 25.38 \pm 0.06$ \\
\Bff &$24.73 \pm 0.06$& $28.11 \pm 0.07 $\\
\Bfr & $23.42 \pm 0.05$ &$26.96 \pm 0.07$ \\
\Bfpp & $2.82 \pm 0.02$& $3.16 \pm 0.02$ \\
\Brpp & $2.64 \pm 0.02 $ & $3.02 \pm 0.02$ \\
\Bpppp & $0.98 \pm 0.01$ & $1.09 \pm 0.01$ \\
    \hline \hline
  \end{tabular}
  \label{t_eps}
\end{table}

\section{Event Model}
\label{Event Model}
The branching fraction is extracted from an extended six-dimensional unbinned maximum likelihood fit to \De, \Fsb, $M_1$, $M_2$, $H_1$ and $H_2$ in the $l^{\rm th}$ $r$-bin and SVD configuration $s$, where $M_k$ and $H_k$ represent the invariant dipion mass \Mpp\ and helicity parameter \cH\ of the $k^{\rm th}$ \rz\ candidate. If linear correlations between fit variables do not cause a noteworthy bias, the probability density function (PDF) for each event $i$, is taken as the product of individual PDFs for each variable ${\cal P}(\De^{i},\Fsb^{i},M_1^{i},M_2^{i},H_1^{i},H_2^{i}) = {\cal P}(\De^{i}) \times {\cal P}(\Fsb^{i}) \times {\cal P}(M_1^{i})\times {\cal P}(M_2^{i}) \times {\cal P}(H_1^{i})\times {\cal P}(H_2^{i})$; otherwise, correlations between the fit variables are taken into account. We consider 17 components in the event model, where most resonances are described by a relativistic Breit-Wigner
\begin{equation}
  BW(m_{\pip\pim})  \equiv \frac{m_{0}\Gamma(m_{\pip\pim})}{(m_{\pip\pim}^{2}-m_{0}^{2})^{2} + m_{0}^{2}\Gamma^{2}(m_{\pip\pim})},
\label{e_bw}
\end{equation}
with a mass-dependent width
\begin{equation}
\Gamma(m_{\pip\pim}) = \Gamma_{0} \biggl(\frac{p_{\pi}}{p_{0}}\biggr)^{3} \biggl(\frac{m_{0}}{\Mpp}\biggr)B_{1}(p_{\pi});
\end{equation}
$p_{\pi}$ is the momentum of a resonance daughter in the resonance frame and $m_{\pip\pim}$ is the invariant mass of a pion pair. $\Gamma_{0}$ and $m_{0}$ are the width and invariant mass of the nominal resonance, such as \rz\ emerging from $B$ decays and $p_{0}$ is the nominal momentum of a pion daughter from a nominal $\rz$. $B_1(p_{\pi}) = \sqrt{\frac{1+(3p_0)^2}{1+(3p_{\pi})^2}}$ is a Blatt-Weisskopf form factor, as described in Ref.~\cite{PDG}.
The PDF for \Fsb\ for all components $j$ are sums of two asymmetric-width (bifurcated) Gaussians in each $r$-bin. The PDF for $H_1$ and $H_2$ for all backgrounds are two-dimensional (2D) histograms, further symmetrized by reflecting the entries along the diagonal, $H_1\leftrightarrow H_2$. We use sets of several histograms in bins of other fit variables if a treatment of the corresponding correlation is required. If smoothing is applicable, e.g. preserves peaks, we use the algorithm 353QH~\cite{TSmooth} for one-dimensional histograms or kernel algorithms else. Chebyshev polynomials are multiplied by a constant factor $c_{\rm order}^{\rm component}$ where the subscript labels the order of the corresponding polynomial and the superscript labels the corresponding component, e.g. nc (cc) for $b\to c$ transitions in neutral (charged) $B$ decays. The different components of the PDF are described below and summarized in Table~\ref{tab_model}: 

\begin{itemize}
\item The signal model shape is determined from correctly reconstructed signal MC events for each polarization. The PDF for \De\ is taken to be a sum of two bifurcated Gaussians. The distribution in the $M_1$-$M_2$ plane is modeled with a product of two relativistic Breit-Wigner functions and the distribution in the $H_1$-$H_2$ plane is described by 
\begin{equation}
\frac{1}{\Gamma}\frac{d^{2}\Gamma}{d\cos\theta_{H1}d\cos\theta_{H2}} = \frac{9}{4}\biggl[\frac{1}{4}(1 - f_L)\sin^{2}\theta_{H1}\sin^{2}\theta_{H2} + f_L\cos^2\theta_{H1}\cos^2\theta_{H2}\biggr],
\label{e_helicity}
\end{equation}
 where $f_L = |A_0|^{2}/\sum |A_{i}|^{2}$ is the fraction of longitudinal polarization; $A_{0}$, $A_{+1}$ and $A_{-1}$ are the helicity amplitudes. We correct the mass and helicity PDFs for the reconstruction efficiency; both, the mass- and the helicity angle-dependent efficiencies, are obtained from fully simulated signal MC events. 

The mis-reconstructed model shape is determined from incorrectly reconstructed signal MC events for each polarization and is described by histograms for all fit variables.

\item The continuum model shape is studied with off-resonance data. Since the off-resonance data contains only a fraction of the number of continuum events expected in this measurement, we keep the entire shape free in the fit to extract the branching fraction. The PDF for \De\ is taken to be a first-order Chebyshev polynomial. The \Mpp\ shape is the sum of a second order polynomial, a Breit-Wigner for $\rz$ and a Breit-Wigner for $f_0(980)$. The helicity angle distribution is described by a histogram and \Fsb\ is modeled as described before; in addition, we account for a correlation with \Mpp by multiplying its mean with the factor $a^{l} = p_0^l(M_1 + M_2) + 1$, where $p_0^l$ is a constant in each $r$-bin.  

\item The PDF for the background due to $B$ decays from $b\to c$ transitions (charm $B$ decays) is determined from a large sample of MC events containing $10$ times the number of expected events and is further divided into a neutral and a charged $B$ sample. For both samples, \De\ is correlated with the helicity angles and therefore its PDF in each sample is formed in different bins. We model the neutral component in four regions of the $H_1$ versus $H_2$ distribution by first-order Chebyshev polynomials where we also add a Gaussian if either a) $|H_{k}|>0.65$ or b) $|H_{j}| >0.65$ and $-0.5<H_{k}<0.65$ with $j\neq k$,
\begin{equation}
 {\cal P}^{\rm charm}_{\BzBzb}(\De)_{H_1,H_2} \equiv (fG(\De, \mu, \sigma) + (1-f)(c_{1}^{\rm nc}C_{1}(\De)))_{H_1,H_2}.
\label{e_gmdE}
\end{equation}
The charged component's \De\ distribution is described by the sum of Chebyshev polynomials up to third order,
\begin{equation}
 {\cal P}^{\rm charm}_{\BpBm}(\De)_{H_1, H_2} \equiv \sum_{i=1,2}c_{i}^{\rm cc}C_{i}(\De) + a(H_1, H_2)c_{3}^{\rm cc}C_{3}(\De).
\label{e_gcdE}
\end{equation}
 where the cubic term is multiplied with a factor $a$, obtained from MC events, that accounts for the correlation with the helicity angles: $a(H_1, H_2) = b^{\rm charm}_{\BpBm}(|H_1| + |H_2|)$ if $|H_{k}|>0.5$ with $b^{\rm charm}_{\BpBm} = 0.68 \pm 0.01$, otherwise $a(H_1, H_2) = 1$.

The \Mpp\ shape of the charm $B$ decays is described by a first order Chebyshev polynomial where a Gaussian is added to the neutral $B$ decays if $|H_{k}| > 0.65$,
\begin{equation}
 {\cal P}^{\rm charm}_{\BzBzb}(\Mpp)_{H_1,H_2} \equiv f_{H_1,H_2} c_{1}C_{1}(\Mpp) (1-f_{H_1,H_2})G(\Mpp, \mu, \sigma).
\label{e_gmM}
\end{equation}
The widths of the Gaussians of both \Fsb\ shapes are multiplied with a constant $d^l$ in the four corners of the $H_1$, $H_2$ plane, a) $|H_{k}|>0.7$ or b) $H_{j}>0.5$ and $H_{k}<-0.5$, $j\neq k$. 

\item The PDF for the background due to $B$ decays from $b\to u,d,s$ transitions (charmless $B$ decays) is determined from a large sample of corresponding MC events containing $50$ times the number of expected events and also divided into a neutral and a charged category. The \De\ distribution of neutral charmless $B$ decays is described by a Gaussian plus a third-order Chebyshev polynomial, and that of charged charmless $B$ decays by a first-order Chebyshev polynomial ${\cal P}^{\rm charmless}_{\BpBm} = c_{cr}C_1(\De)$, where $c_{cr} \equiv a^{\rm charmless}_{\BpBm}(M_1+M_2) +  b^{\rm charmless}_{\BpBm}(M_1 \times M_2)$ with $ a^{\rm charmless}_{\BpBm} =  -2.44 \pm 0.47$ and $b^{\rm charmless}_{\BpBm} = 5.46 \pm 1.08$ accounts for the correlation with the dipion masses. Since the neutral charmless $M_1$, $M_2$ distribution is correlated with $H_1$, $H_2$, we describe its shape by the sums of Gaussians and Chebyshev polynomials in five bins of $H_1$, $H_2$; in addition, in most bins, a correlation between the two $\pip\pim$ masses had to be taken into account. The charged \Mpp\ distribution is modeled by a smoothed 2D histogram.

\item The PDF shapes for the remaining four-pion states ($B^0 \to \pip\pim\pip\pim, \rho^0\pip\pim, f_0\rz,$ $f_0f_0,f_0\pip\pim, a_1^{\pm}\pi^{\mp},  a_2^{\pm}\pi^{\mp}$ and $ b_1^{\pm}\pi^{\mp}$) are determined from individually generated MC samples. For the decay \Brpp, we assume a phase space distribution and account for this assumption in the systematic uncertainty. \De\ is described in a similar manner to the signal component. Since we include mis-reconstructed events in the model, a polynomial is added to their \De\ PDFs. For the non-resonant decay, the correlation of \De\ with the helicity angles is incorporated by taking a different width of the \De's core Gaussian for the center of the $H_1$-$H_2$ plane. The $M_1$, $M_2$ shapes are modeled by 2D histograms, except for \Bff\ and \Bfr, where products of two Breit-Wigners are used. For \Bfr, the correlation between $M_1$ and $M_2$ is taken into account. The mean of the Gaussian of the $B\to a_1^{\pm}\pi^{\mp}$ \Fsb\ shape is multiplied with $a=1+b_{a_1\pi}(M_1+M_2)$, to account for the correlation with the masses.
\end{itemize}
For \De\ and \Fsb, we incorporate calibration factors that correct for the difference between data and MC by calibrating the mean and width of the core bifurcated Gaussians. They are determined from a large-statistics control sample \dpi\ and are used for the \De\ PDFs of all four-pion final states and for the \Fsb\ PDFs of all $B\bar{B}$ modes, whose shapes are all determined from MC events. Furthermore, the signal's core Gaussian is made common among all four-charged-pion final states for \De\ and among all $B\bar{B}$ modes for \Fsb. 

\begin{table}
  \centering
  \caption{Summary of the event model. The signal PDF for longitudinally (transversely) polarized $\rz$s is denoted by the superscript $LP$ ($TP$) and the subscript $true$ ($mr$) implies PDFs for correctly (mis-) reconstructed signal events. G denotes a Gaussian and dBG denotes the sum of two bifurcated Gaussian, where the superscript sig indicates that the core Gaussian is inherited from the signal PDF and the subscript r indicates a separate description in each $r$-bin. $C_i$ are sums of Chebyshev polynomial of different order, where the subscript labels the number of used orders; e.g. $C_{1-3} = C_1+C_2+C_3$ and $C_{2,4} = C_2+C_4$. A relativistic Breit-Wigner is denoted by BW, the subscript labels the resonance. Histograms are denoted by H, where the subscript sm indicates smoothing and the superscript the dimension. Correlations with other variables, if taken into account, are denoted with $\rm PDF|_{\rm variable(s)}$. Since the two dipion masses and helicity angle PDFs are symmetric under $\rz_1 \leftrightarrow \rz_2$ only the PDF for one \rz\ is given, the actual PDF is then the product of two PDFs, one for each \Mpp(\cH). An asymmetry within the two-dimensional \Mpp\ PDF is denoted as $PDF|^{M}$.}
  \begin{tabular}
    {@{\hspace{0.25cm}}c@{\hspace{0.25cm}}    @{\hspace{0.25cm}}c@{\hspace{0.25cm}} @{\hspace{0.25cm}}c@{\hspace{0.25cm}}  @{\hspace{0.25cm}}c@{\hspace{0.25cm}}  @{\hspace{0.25cm}}c@{\hspace{0.25cm}}}
    \hline \hline
    Component &  \De\  & \Mpp & \cH\ & \Fsb\ \\
    \hline
 $B\to\rz\rz|_{true}^{LP}$& dBG & $BW_{\rz}$ & Eq.~\ref{e_helicity} & $\rm dBG_{r}$ \\
  $B\to\rz\rz|_{true}^{TP}$& dBG & $BW_{\rz}$ & Eq.~\ref{e_helicity} & $\rm dBG_{r}$ \\
  $B\to\rz\rz|_{mr}^{LP}$ & $\rm H_{sm}^{1D}$ & $\rm H_{sm}^{2}$  & $\rm H^{2D}$ & $\rm dBG^{sig}_{r}$ \\
  $B\to\rz\rz|_{mr}^{TP}$& $\rm H_{sm}^{1D}$ & $\rm H_{sm}^{2}$  & $\rm H^{2D}$ & $\rm dBG^{sig}_{r}$ \\
 \hline
continuum & $C_1$ & $BW_{\rz}+BW_{f_0}+C_{2,4}$ & $\rm H^{2D}$  & $\rm dBG_{r}|_{M1, M2}$ \\
 \hline
$B(\bar{B})\to$ charm &  $C_1+G|_{H1,H2}$ &$C_1 + G|_{H1,H2}$ &$\rm H^{2D}|_{\De}$ & $\rm dBG^{sig}_{r}|_{H1,H2}$ \\
$B^\pm\to$ charm & $C_{1-3}|_{H1,H2}$&$C_{1-3}$& $\rm H^{2D}|_{\Fsb}$ & $\rm dBG^{sig}_{r}|_{H1,H2}$ \\
$B(\bar{B})\to$ charmless & $C_3 + G$ & $(C_i + G)|_{H1,H2,}^{M}$ & $\rm H^{2D}|_{\De}$ &  $\rm dBG^{sig}_{r}$\\
$B^\pm\to$ charmless &$C_1|_{M1,M2}$ & $\rm H_{sm}$ & $\rm H^{2D}$ & $\rm dBG^{sig}_{r}$\\
 \hline
$B\to\pip\pim\pip\pim$ & $\rm dBG^{sig}|_{H1,H2}$ &  $\rm H_{sm}^{2D}$ & $\rm H^{2D}$ & $\rm dBG^{sig}_{r}$\\
$B\to\rz\pip\pim$ & $\rm dBG^{sig}$ & $\rm H^{2D}$ & $\rm H^{2D}$ & $\rm dBG^{sig}_{r}$\\
$B\to f_0\pip\pim$ &  $\rm dBG^{sig}$& $\rm H^{2D}$ & $\rm H^{2D}$ & $\rm dBG^{sig}_{r}$\\
$B\to f_0f_0$ & $\rm dBG^{sig}$ & $BW_{f_0}$ & $\rm H^{2D}$ & $\rm dBG^{sig}_{r}$\\
$B\to f_0\rz$ & $\rm dBG^{sig}$ & $BW_{\rz}\cdot BW_{f_0}|^{M}$& $\rm H^{2D}$ & $\rm dBG^{sig}_{r}$\\
$B\to a_1^\pm\pi^\mp$ & $\rm dBG^{sig}$ &  $\rm H_{sm}^{2D}$ & $\rm H^{2D}$ & $\rm dBG^{sig}_{r}|_{M1,M2}$\\
$B\to a_2^\pm\pi^\mp$ & $\rm dBG^{sig}$ &  $\rm H_{sm}^{2D}$ & $\rm H^{2D}$ & $\rm dBG^{sig}_{r}$\\
$B\to b_1^\pm\pi^\mp$ & $\rm dBG^{sig}$ &  $\rm H_{sm}^{2D}$ & $\rm H^{2D}$ & $\rm dBG^{sig}_{r}$\\
     \hline \hline
  \end{tabular}
  \label{tab_model}
\end{table}

Besides the dominant four-pion contribution \aonepi, we also fix the other quasi-two-body $B^0\to X^{\pm}\pi^{\mp}$ decay modes as listed in Table~\ref{tab_pb}, where we assume the unknown branching fraction of \atwopi\ to be $10\%$ of \aonepi's. A recent measurement by Belle supports this assumption~\cite{jeremy_a1pi}. The other two branching fractions are assumed to take their current world average values, given in Ref.~\cite{PDG}. Throughout this paper, for the modes including an $f_0$, the exclusive branching fraction for $f_0\to \pip\pim$ is incorporated; e.g., ${\cal B}(B^{0}\rightarrow f_0 f_0)$ stands for ${\cal B}(B^{0}\rightarrow f_0 f_0)\times{\cal B}( f_0\to \pip\pim)^2$. 

\begin{table}
  \centering
  \caption{List of peaking backgrounds, assumed branching fractions and their expected yields $N^{\rm s}_{\rm expected}$ for the two detector configurations; SVD1 and SVD2.}
  \begin{tabular}
    {@{\hspace{0.5cm}}c@{\hspace{0.25cm}}  @{\hspace{0.25cm}}c@{\hspace{0.25cm}}  @{\hspace{0.25cm}}c@{\hspace{0.25cm}}  @{\hspace{0.25cm}}c@{\hspace{0.25cm}}  @{\hspace{0.25cm}}c@{\hspace{0.5cm}}}
    \hline \hline
    Mode & ${\cal B}$ $(\times 10^{-6})$ & $N^{\rm SVD1}_{\rm expected}$ & $N^{\rm SVD2}_{\rm expected}$ \\
    \hline
  $\Bz \rightarrow \aone [\pi^{\pm} \pi^{\mp} \pi^{\pm}] \pi^{\mp}$ &  $16.5 \pm 2.5$&  $65$ & $299$\\
  $\Bz \rightarrow \atwo [\pi^{\pm} \pi^{\mp} \pi^{\pm}] \pi^{\mp}$ & $1.65 \pm 1.65$ & $1$ & $7$\\
    $\Bz \rightarrow \bone [\pi^{\pm} \pi^{\mp} \pi^{\pm}] \pi^{\mp}$ & $0.17\pm 1.53$ & $1$ & $3$\\
     \hline \hline
  \end{tabular}
  \label{tab_pb}
\end{table}

The total likelihood for $116081$ signal candidate events in the fit region is
\begin{equation}
  {\cal L} \equiv \prod_{l,s} \frac{e^{-\sum_{j}N^{s}_{j}\sum_{l,s}f^{l,s}_{j}}}{N_{l,s}!} \prod^{N_{l,s}}_{i=1} \sum_{j}N^{s}_{j}f^{l,s}_{j}{\cal P}^{l,s}_{j}(\De^{i},\Fsb^{i},M_1, M_2, H_1, H_2),
\end{equation}
which runs over event $i$, component $j$, $r$-bin $l$ and SVD configuration $s$. Instead of two free signal yields $N^{s}_{\rm Sig}$ for each detector configuration, branching fractions for the four-pion final states ($j$) are chosen as single free parameters ${\cal B}(B\to X)$ and incorporated into the fit with
\begin{equation}
  N^{s}_{j} = {\cal B}(B^{0}\to f)N^{s}_{\BBbar}\epsilon^{s}_{j}\eta^{s}_{j},
\label{e_N}
\end{equation}
where $\epsilon^{s}_{j}$ and $\eta$ are the efficiencies and correction factors described in Sec.~\ref{Event Selection}. Because of the two possible polarizations in \Brr\ decays, Eq.~\ref{e_N} takes the distinct forms; for example, for longitudinally polarized \rz s (LP):
\begin{equation}
N_{\rm LP}^{s}  = {\cal B}(\Brr)f_{L}N_{\BBbar}^{s}\epsilon_{\rm LP}^{s}\eta_{\rm LP}^{s},
\end{equation}
and similarly with $(1-f_L)$ replacing $f_L$ for transverse polarization. The fraction of events in each $r$-bin $l$, for component $j$, is denoted by $f^{l,s}_{j}$. The fractions of all $B\bar{B}$ components, $f^{l,s}_{\rm B\bar{B}}$, have been calibrated with the \dpi\ control sample (see Sec.~\ref{Validity Checks and Significance}).
In the fit to data, we also float $f_L$, the yields $N^{s}_{\qqbar}$, $N^{{\rm charm}, s}_{\BzBzb}$ and $N^{{\rm charmless}, s}_{\BzBzb}$ and the parameters of the shape of the continuum model. The remaining yields are fixed to the values given in Table~\ref{tab_bf_fixed} as determined from MC simulation.
\begin{table}
  \centering
  \caption{Summary of yields fixed relative to other yields free in the fit for the two detector configurations. The mis-reconstructed yield is fixed relative to the signal yield; charmed and charmless \BpBm\ background yields are fixed relative to their respective \BzBzb\ background yields. The central values are obtained from MC simulation; the errors are statistical.}
  \begin{tabular}
    {@{\hspace{0.5cm}}c@{\hspace{0.25cm}}  @{\hspace{0.25cm}}c@{\hspace{0.25cm}}  @{\hspace{0.25cm}}c@{\hspace{0.5cm}}}
    \hline \hline
    Component & Yield SVD1 & Yield SVD2\\
    \hline
    $N^{s}_{\rm Mis}$ & $(0.09 \pm 0.0006)N^{\rm SVD1}_{\rm Sig}$ & $(0.09 \pm  0.0006)N^{\rm SVD2}_{\rm Sig}$\\[5pt]
    $N^{{\rm charm},s}_{\BpBm}$ & $(1.40 \pm 0.07)N^{\rm charm, SVD1}_{\BzBzb}$ & $(1.42 \pm 0.04)N^{\rm charm, SVD2}_{\BzBzb}$\\[5pt]
    $N^{{\rm charmless},s}_{\BpBm}$ & $(0.83 \pm 0.08)N^{\rm charmless, SVD1}_{\BzBzb}$ & $(0.86 \pm 0.04)N^{\rm charmless, SVD2}_{\BzBzb}$\\[5pt]
    \hline \hline
  \end{tabular}
  \label{tab_bf_fixed}
\end{table}

\section{Fit Result}
\label{Fit Result}
\begin{figure}[h]
  \centering
  \includegraphics[height=150pt,width=!]{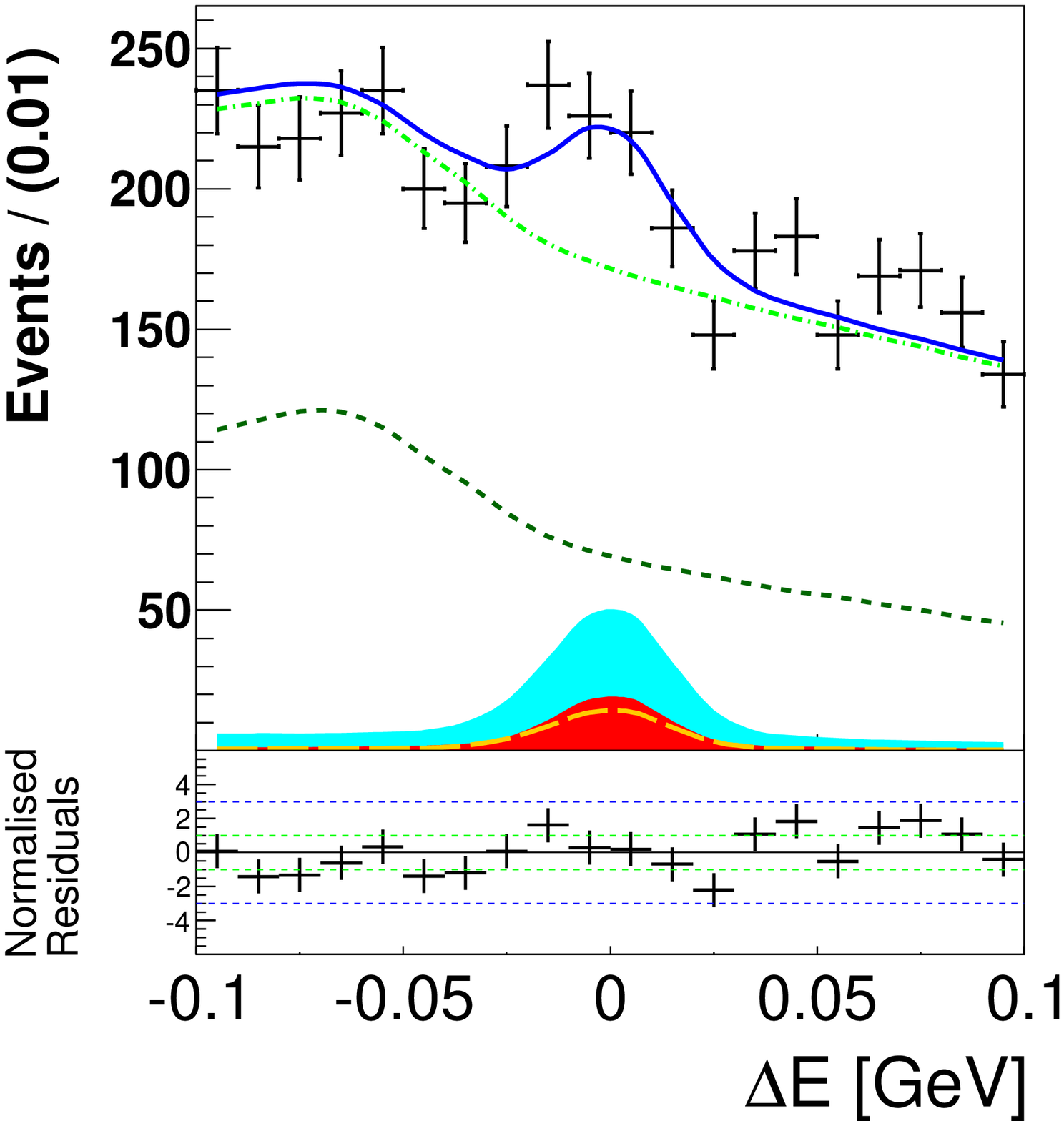}
   \includegraphics[height=150pt,width=!]{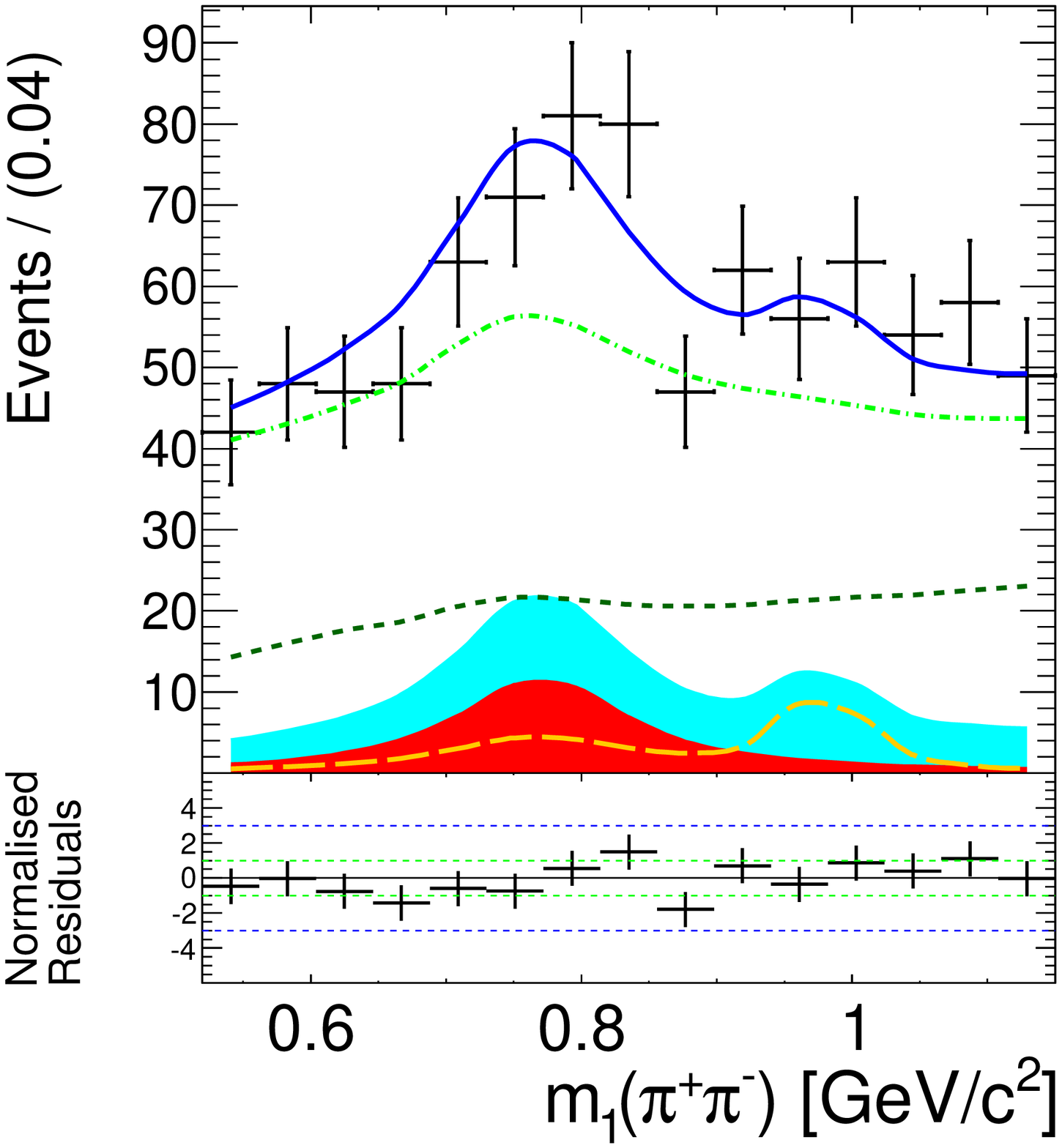}
  \includegraphics[height=150pt,width=!]{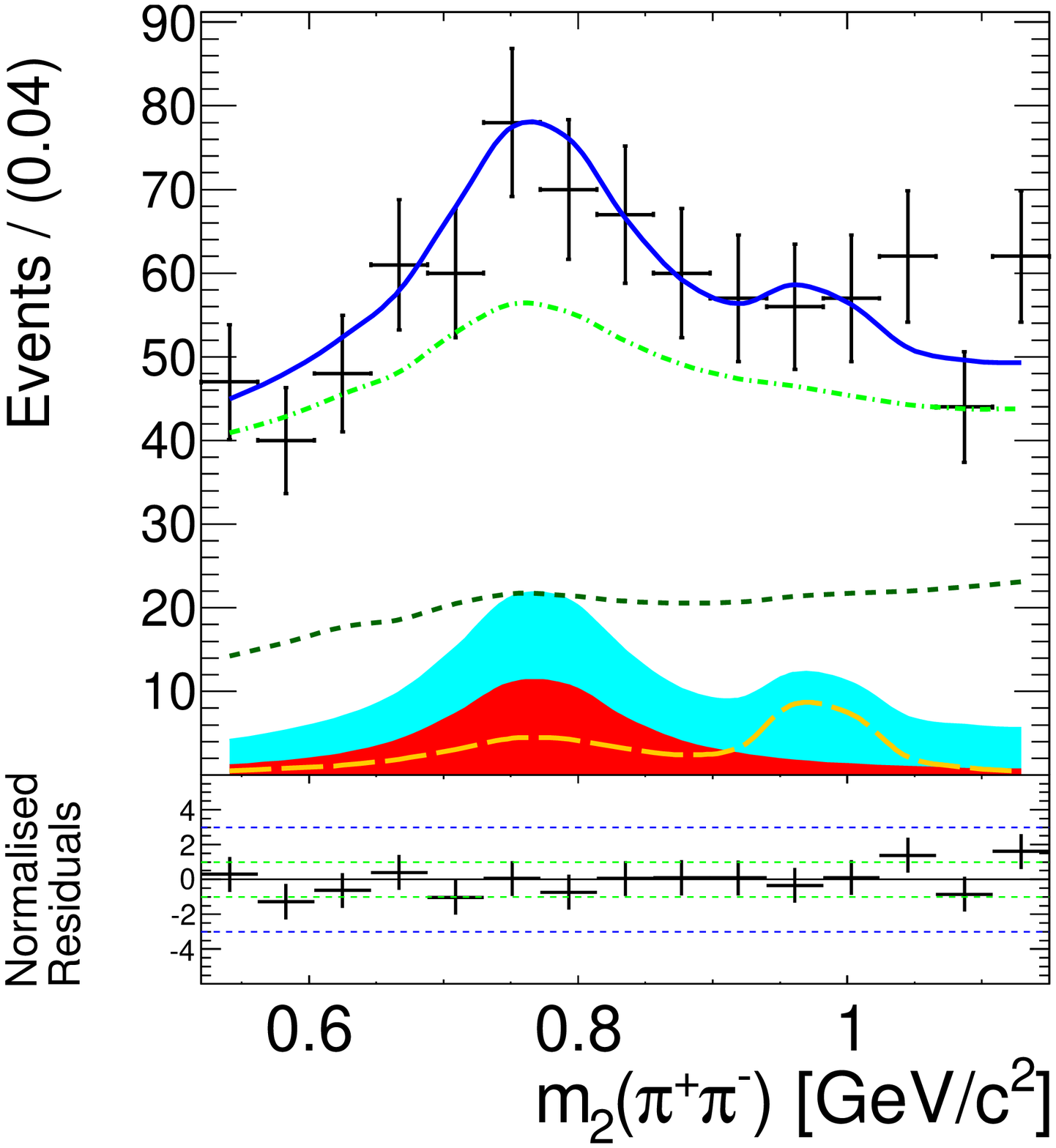}
\put(-350,130){(a)}
  \put(-195,130){(b)}
 \put(-35,130){(c)}

\includegraphics[height=150pt,width=!]{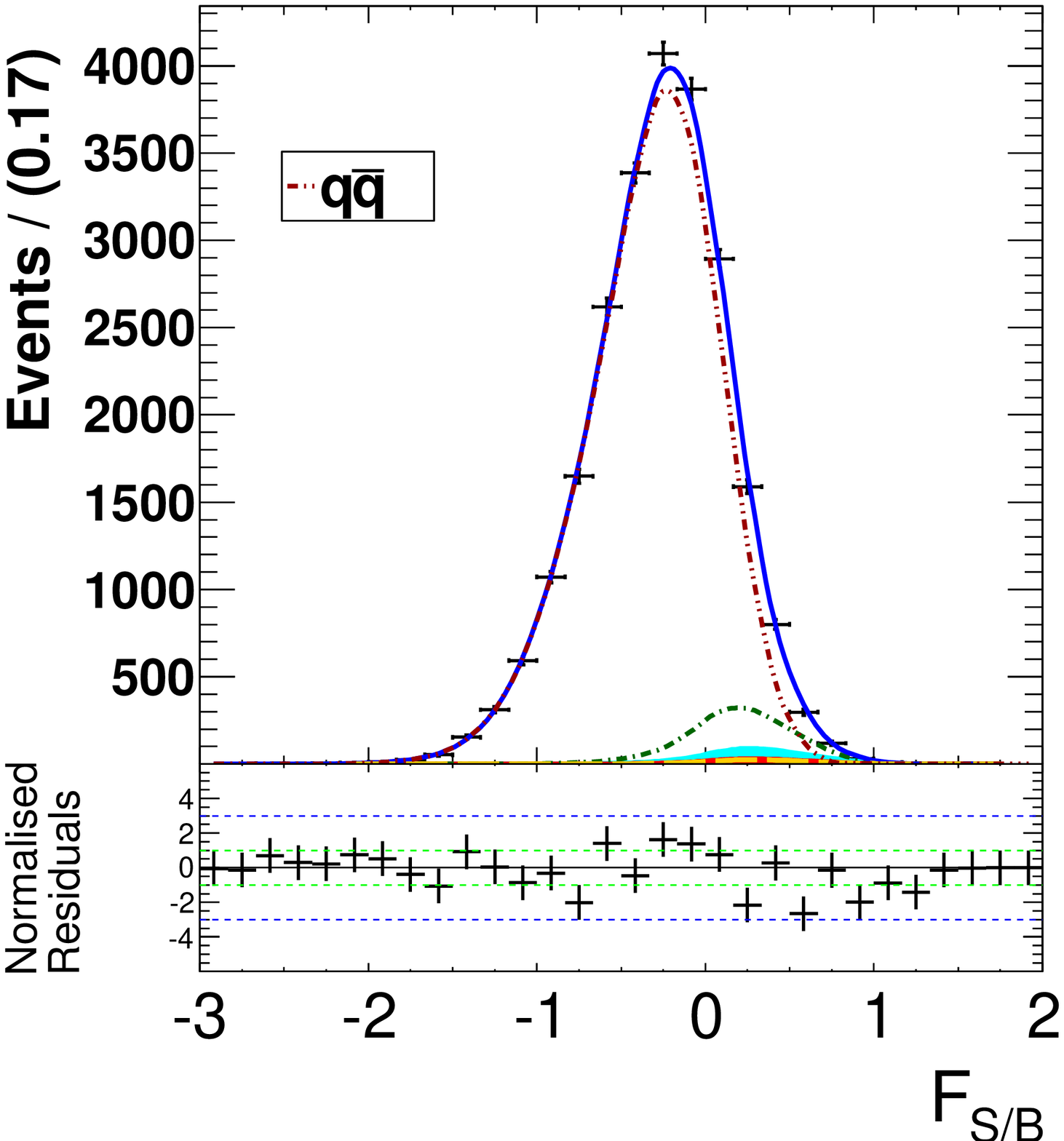}
 \includegraphics[height=150pt,width=!]{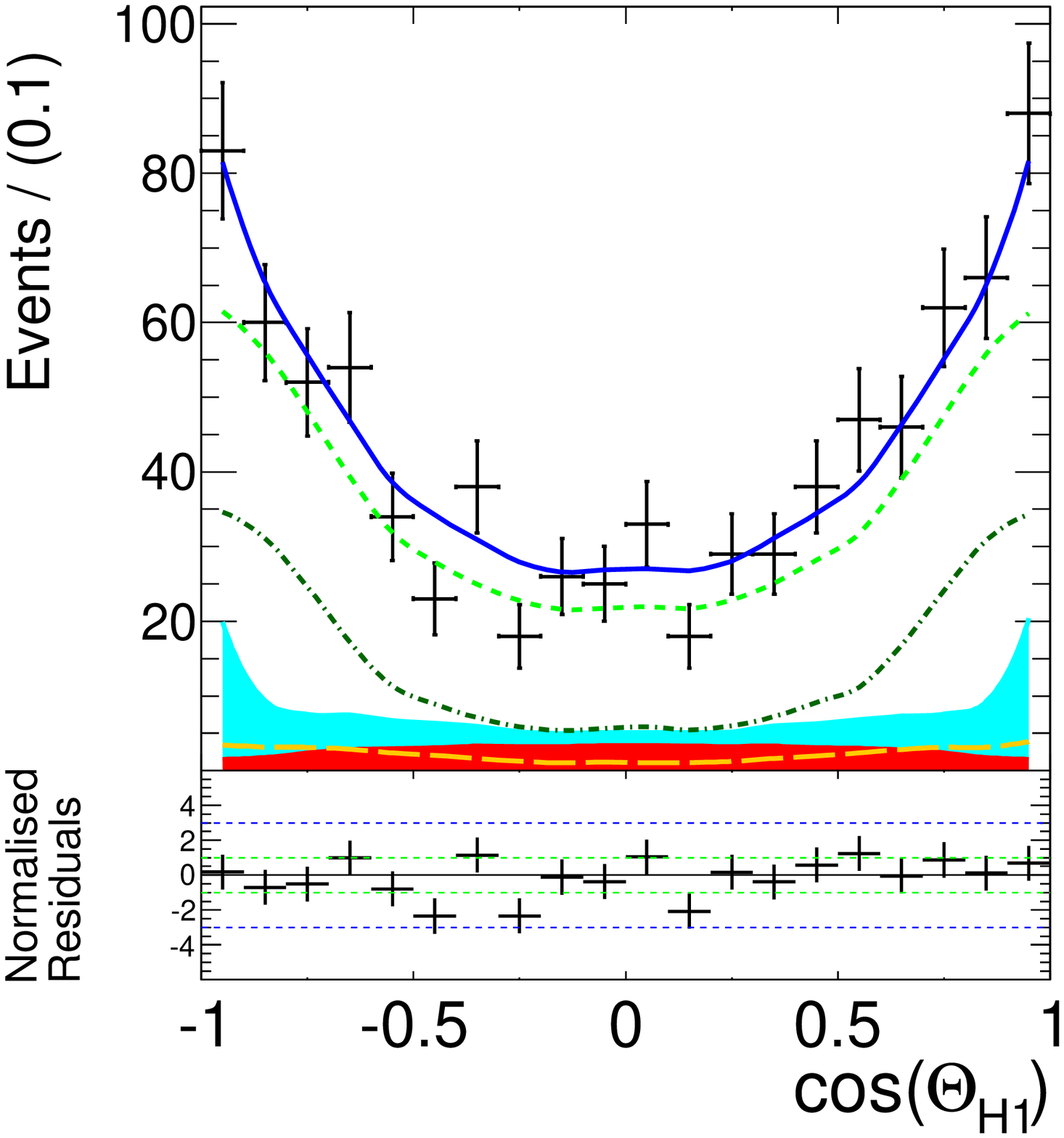}
 \includegraphics[height=150pt,width=!]{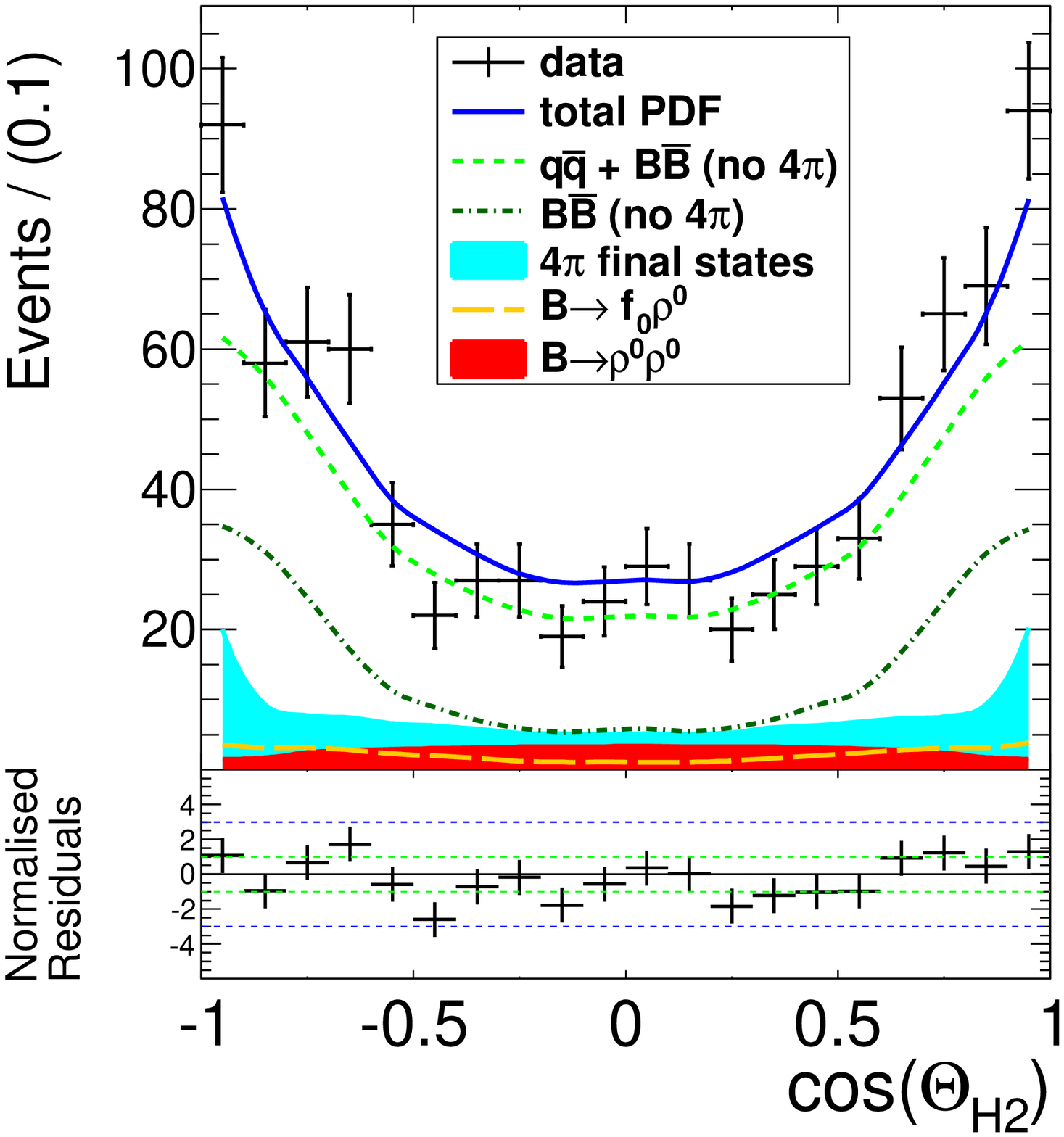}
 \put(-350,130){(d)}
 \put(-195,130){(e)}
 \put(-35,130){(f)}
 \caption{(color online) Projections of the fit to the data. The points represent the data and the solid curves represent the fit result. The shaded (red) areas show the \Brr\ and the (orange) long dashed curves the \Bfr\ contribution. The bright-shaded (cyan) areas show all four-pion final states, the (dark green) dashed curves show the non-peaking $B\bar{B}$ contribution and the (bright green) dash-dotted curves show the total non-peaking  background. Panel (a) shows the \De\ projection in the $\Fsb > 0.5$ region; (d) shows the \Fsb\ projection in the $|\De| < 0.02 \; {\rm GeV}$ region. Here, the (dark red) dash double-dotted curve shows only the continuum background contribution. Panels (b,c) show the \Mpp\ projections and (e,f) the \cH\ projections in the $\Fsb > 0.5$ and $|\De| < 0.02 \; {\rm GeV}$  region. The residuals are plotted below each distribution.}
   \label{fig_bf_data}
\end{figure}

We perform a six-dimensional fit to the data with $90$ free parameters. The projections of the fit results onto \De, $M_1$, $M_2$, $H_1$, $H_2$ and \Fsb\ are shown in Fig.~\ref{fig_bf_data}. Although a clear four-pion final state peak can be seen in the $\De$ distribution, the strongly signal-enhanced plots still demonstrate the dominance of the background, especially continuum, in the projection onto $\Fsb$. The obtained branching fractions, their corresponding yield and upper limits at $90\%$ confidence level are given in Table~\ref{t_fitresult}, together with $f_L$ of \Brr. The statistical correlation coefficients between the observables are given in Table~\ref{t_correlM}.
\begin{table}[h]                                                                                                                                                               
\centering                                                                                                                                                                      
\caption{Branching fractions {$\cal B$} with their corresponding yield, upper limits (UL) at $90\%$ confidence level and the significances (${\cal S}$) as described in Sect.~\ref{Validity Checks and Significance} for modes with a positive yield from the fit to data. Furthermore, the longitudinal polarization fraction $f_L$ of \Brr\ is given. The first and second errors are statistical and systematic, respectively; both errors are also included in the upper limits and significances. ${\cal B}$ stands for the product of all branching fractions involved in the $B$ decay to the four-pion final state.}
  \begin{tabular}                                                                                                                                                               
      {@{\hspace{0.25cm}}l@{\hspace{0.25cm}} @{\hspace{0.25cm}}l@{\hspace{0.25cm}} @{\hspace{0.25cm}}l@{\hspace{0.25cm}} @{\hspace{0.25cm}}l@{\hspace{0.25cm}} @{\hspace{0.25cm}}l@{\hspace{0.25cm}} @{\hspace{0.25cm}}l@{\hspace{0.25cm}} }          
      \hline \hline                                 
Mode & ${\cal B}$ ($10^{-6}$) & Yield & $f_L$ & UL $(10^{-6})$ & ${\cal S}$ ($\sigma$)\\
\hline
$\Brr$ & $ 1.02\pm 0.30  \pm 0.15 $ & $166\pm49$ &  $0.21^{+0.18}_{-0.22} \pm 0.15 $ & n.a. & $3.4$\\
$B^{0}\rightarrow \pip\pim\pip\pim$ & $ -3.58^{+7.75}_{-7.19} \pm 2.10 $ & $-25\pm54 $ & n.a. &$11.2$ & n.a.\\
 $B^{0}\rightarrow \rz\pip\pim$ & $ 1.70^{+4.21}_{-4.12} \pm 5.11 $& $33\pm82 $ & n.a.&$12.0$ & n.a.\\ 
 $B^{0}\rightarrow f_0\pip\pim $ & $ -1.34^{+2.12}_{-1.97} \pm 0.80$& $-27\pm44 $& n.a.&$3.0$ & n.a.  \\ 
 $B^{0}\rightarrow f_0\rz$ & $ 0.78 \pm 0.22 \pm 0.11 $ & $125 \pm 41 $ & n.a.&n.a. & $3.1$\\
 $B^{0}\rightarrow f_0 f_0$ & $ -0.03^{+0.10}_{-0.09} \pm 0.03 $ & $-5\pm17$ & n.a. &$0.2$ & n.a.\\
\hline  \hline
 \end{tabular}                                                                                                                                                                 
\label{t_fitresult}         
 \end{table}

\begin{table}[h]                                                                                                                                                               
\centering     
   \caption{Statistical correlations between the observables.}                                                                                                                                                           
  \begin{tabular}                                                                                                                                                               
      {@{\hspace{0.25cm}}c@{\hspace{0.25cm}} @{\hspace{0.25cm}}c@{\hspace{0.25cm}} @{\hspace{0.25cm}}c@{\hspace{0.25cm}}@{\hspace{0.25cm}}c@{\hspace{0.25cm}} @{\hspace{0.25cm}}c@{\hspace{0.25cm}}@{\hspace{0.25cm}}c@{\hspace{0.25cm}} @{\hspace{0.25cm}}c@{\hspace{0.25cm}}@{\hspace{0.25cm}}c@{\hspace{0.25cm}}}                                                                                                   \hline \hline                                 
${\cal B}(B^{0}\rightarrow X)$ & $\rz\rz$ & $f_L$ & $\pip\pim\pip\pim$ & $\rz\pip\pim$ &  $f_0\pip\pim$ & $f_0f_0$ & $f_0\rz$\\

$\rz\rz$ & $1$ & $0.45  $ & $ -0.03 $ & $ -0.34  $ & $ 0.04 $ & $ 0.02 $ & $-0.02$   \\

$f_L$ & & $1 $ & $0.06  $ & $ 0.00 $ & $0.03  $ & $0.01  $ & $ -0.08$ \\

$\pip\pim\pip\pim$ &  &  &$1$  & $-0.35 $ & $-0.23 $ & $-0.04  $ & $ 0.00$ \\

$\rz\pip\pim$  &  &  &  & $1 $ & $-0.07  $ & $0.02 $ & $ -0.24 $\\

 $f_0\pip\pim$  &  &  &  &  &$1 $ & $-0.36 $ & $ -0.20 $\\

$f_0f_0$  &  &  &  &  &  & $1 $ & $0.07$\\

$f_0\rz$  &  &  &  &  &  & &  $1$\\
\hline
 \hline
 \end{tabular}                                                                                                                                                                 
\label{t_correlM}                                                                                                                                                          
   \end{table}

Ignoring uncertainties and interference effects for the moment, the relative contributions of the components modeled in this analysis are found to be $0.1\%$ $B^{0}\to\rz\rz$,  $0.1\%$ $B^{0}\to f_0\rz$, $92.8\%$ continuum, $6.7\%$ $B\bar{B}$ background and $0.3\%$ for the remaining four-pion final states. 

\begin{figure}[h]                                                                                                                                                               
\centering
\includegraphics[height=!,width=0.32\columnwidth]{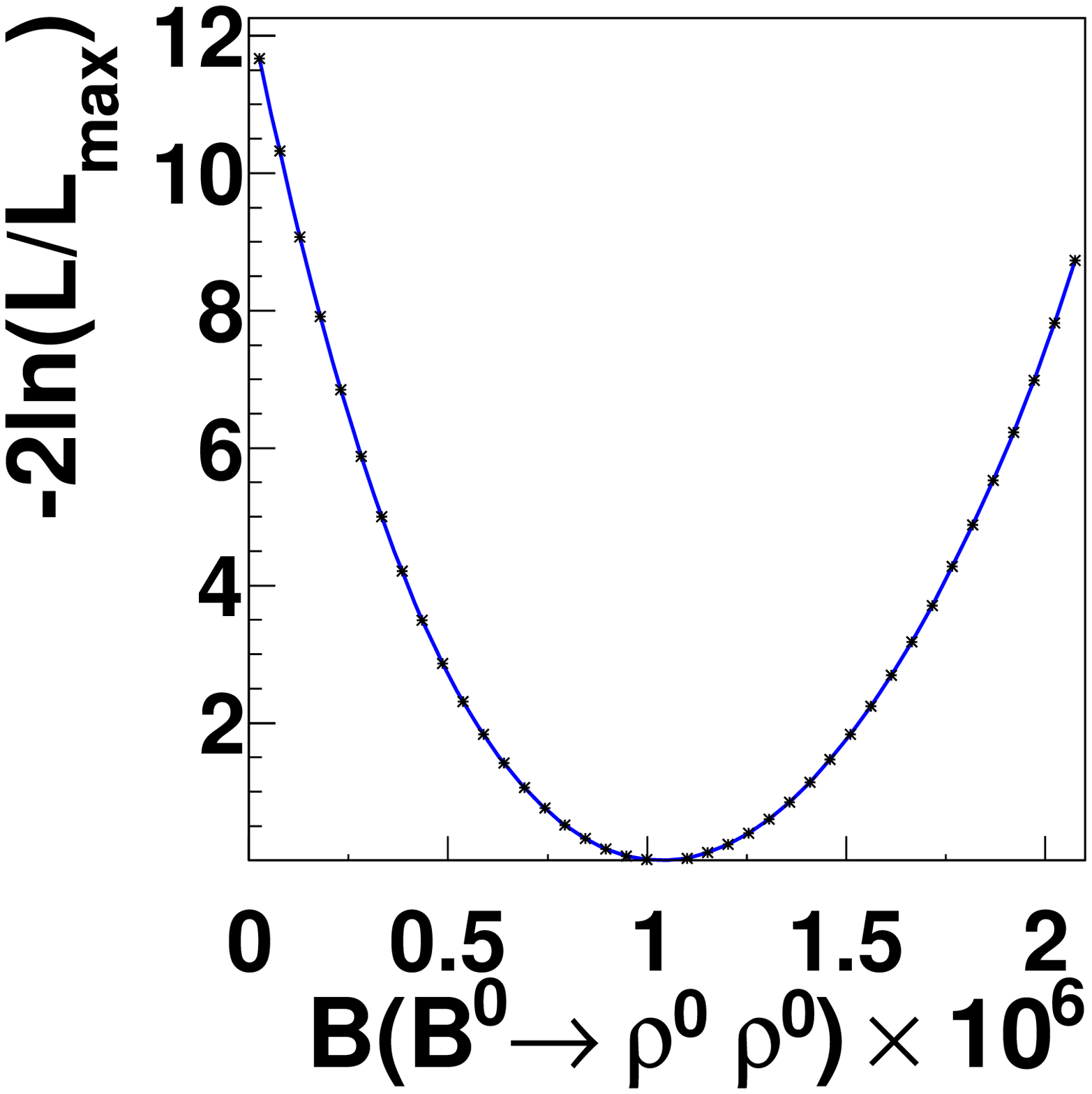}   
      \includegraphics[height=!,width=0.32\columnwidth]{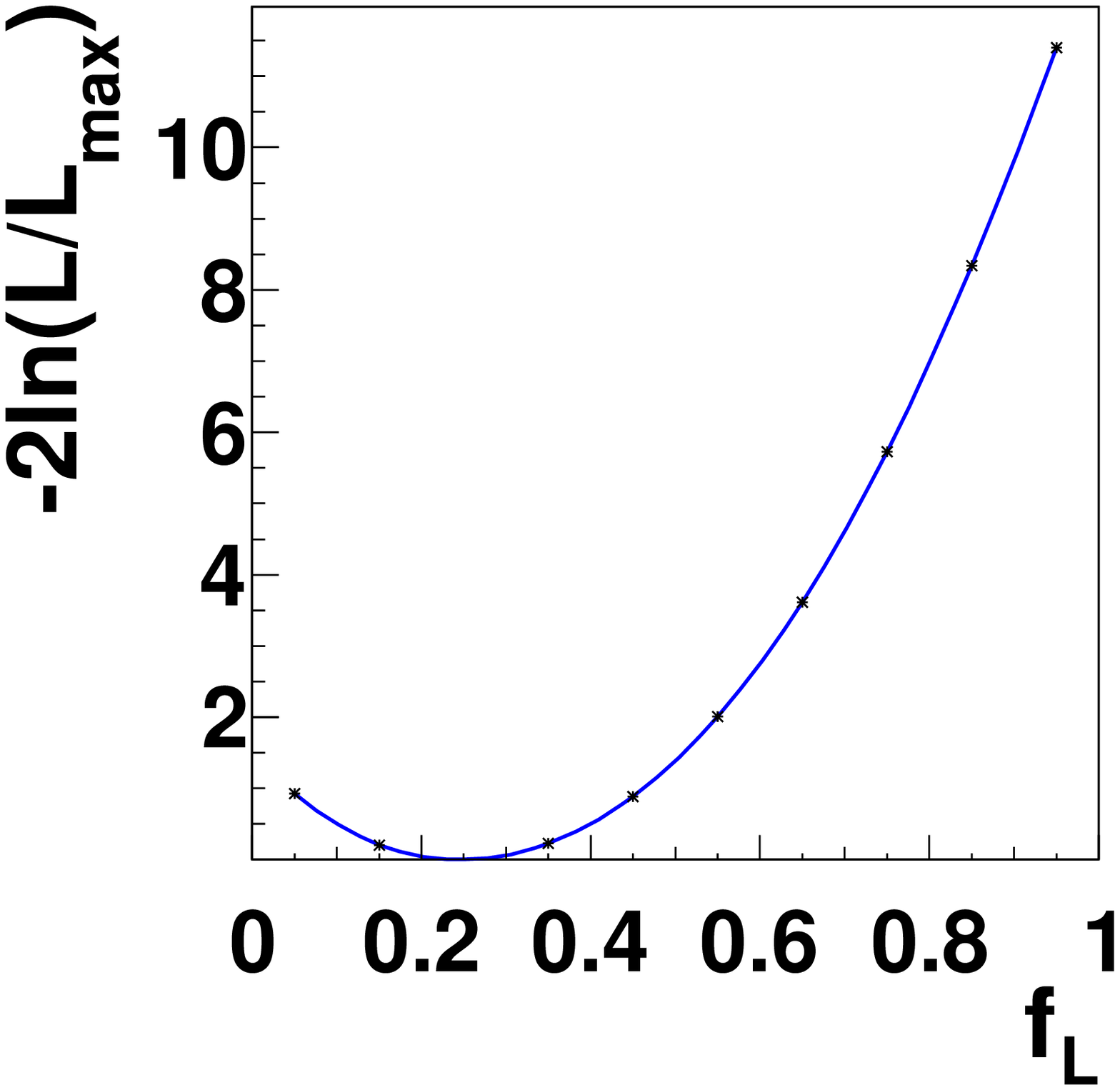}                       
\includegraphics[height=!,width=0.32\columnwidth]{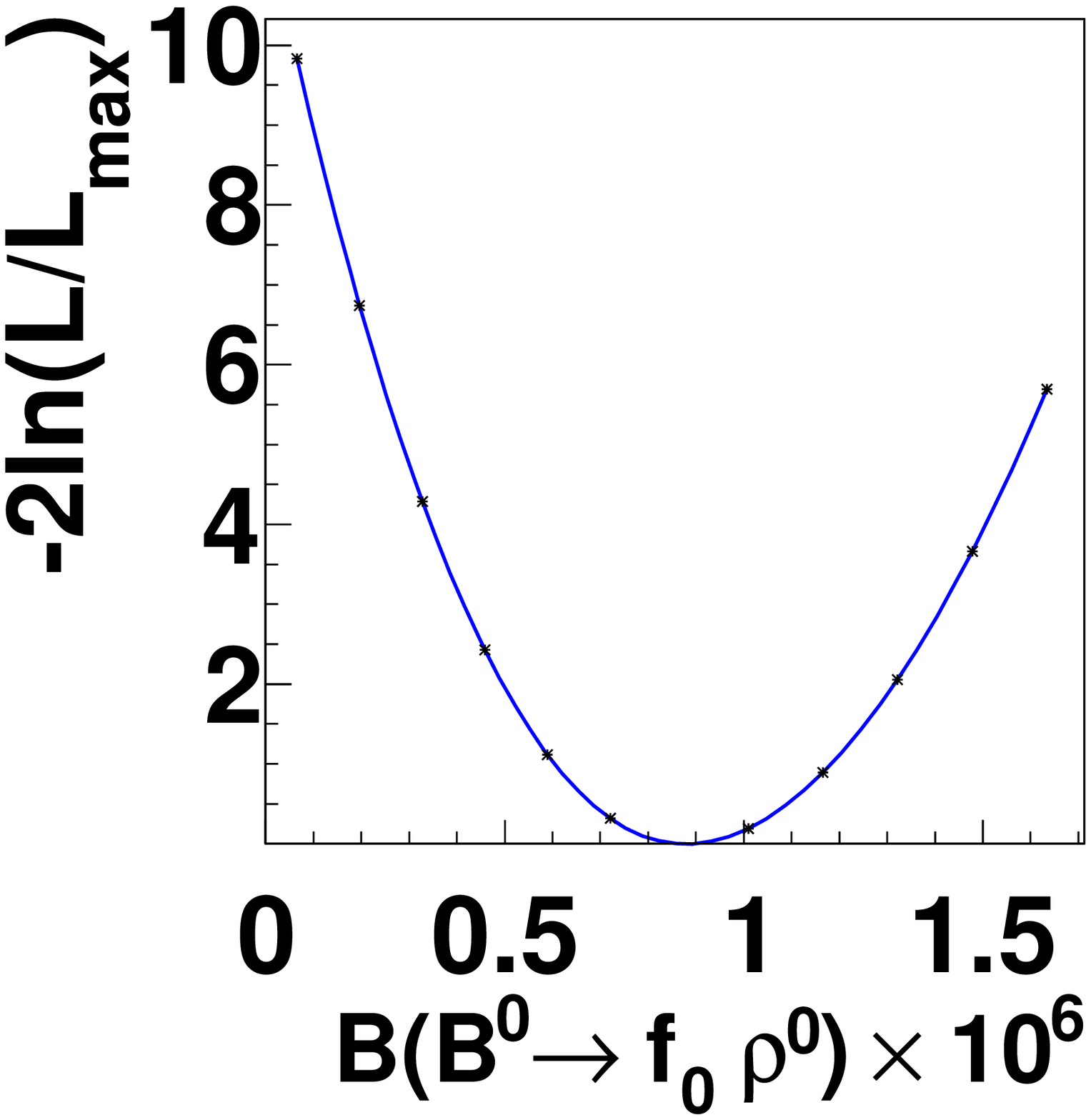}  
        \includegraphics[height=!,width=0.32\columnwidth]{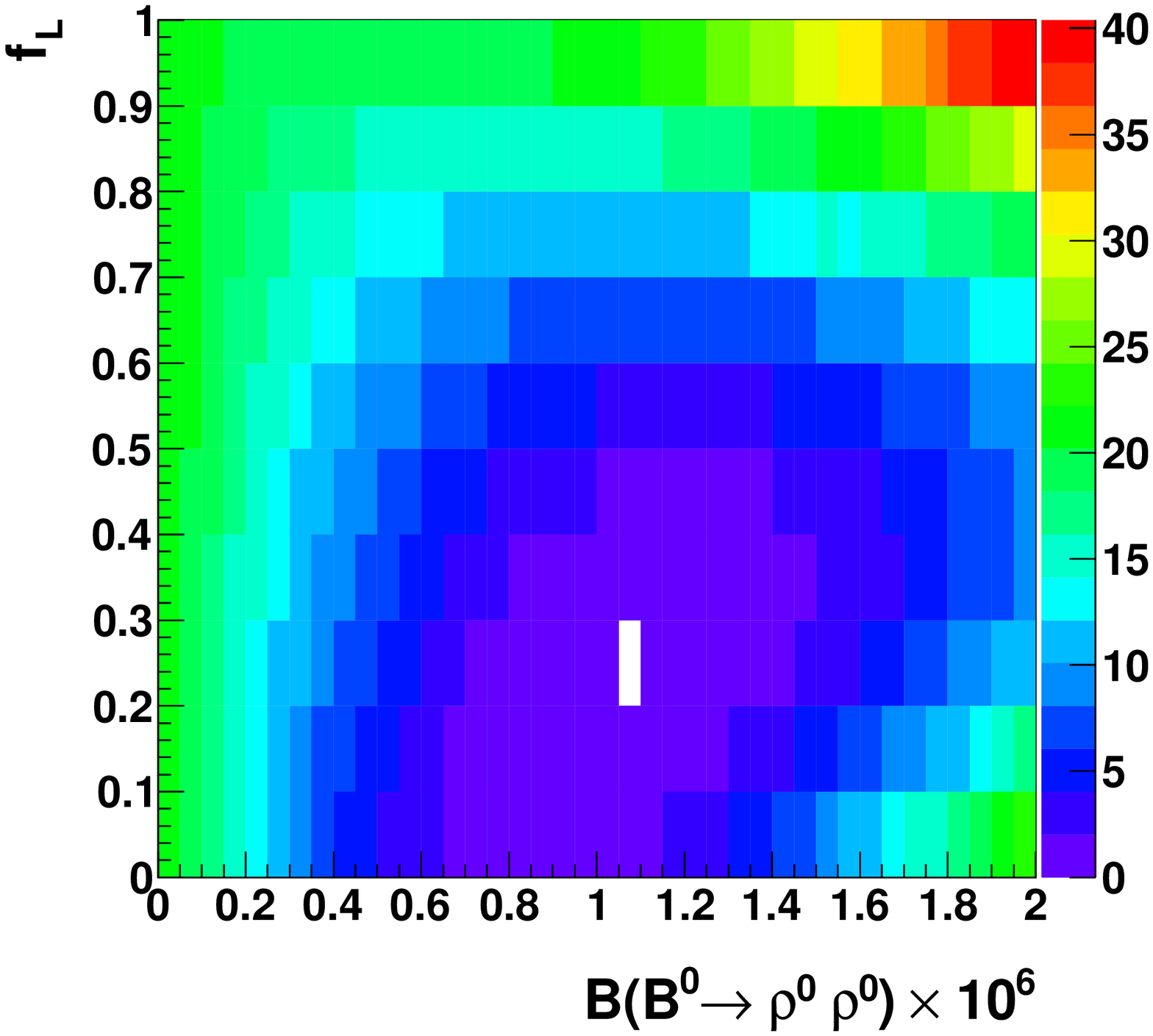}                                                  
        \includegraphics[height=!,width=0.32\columnwidth]{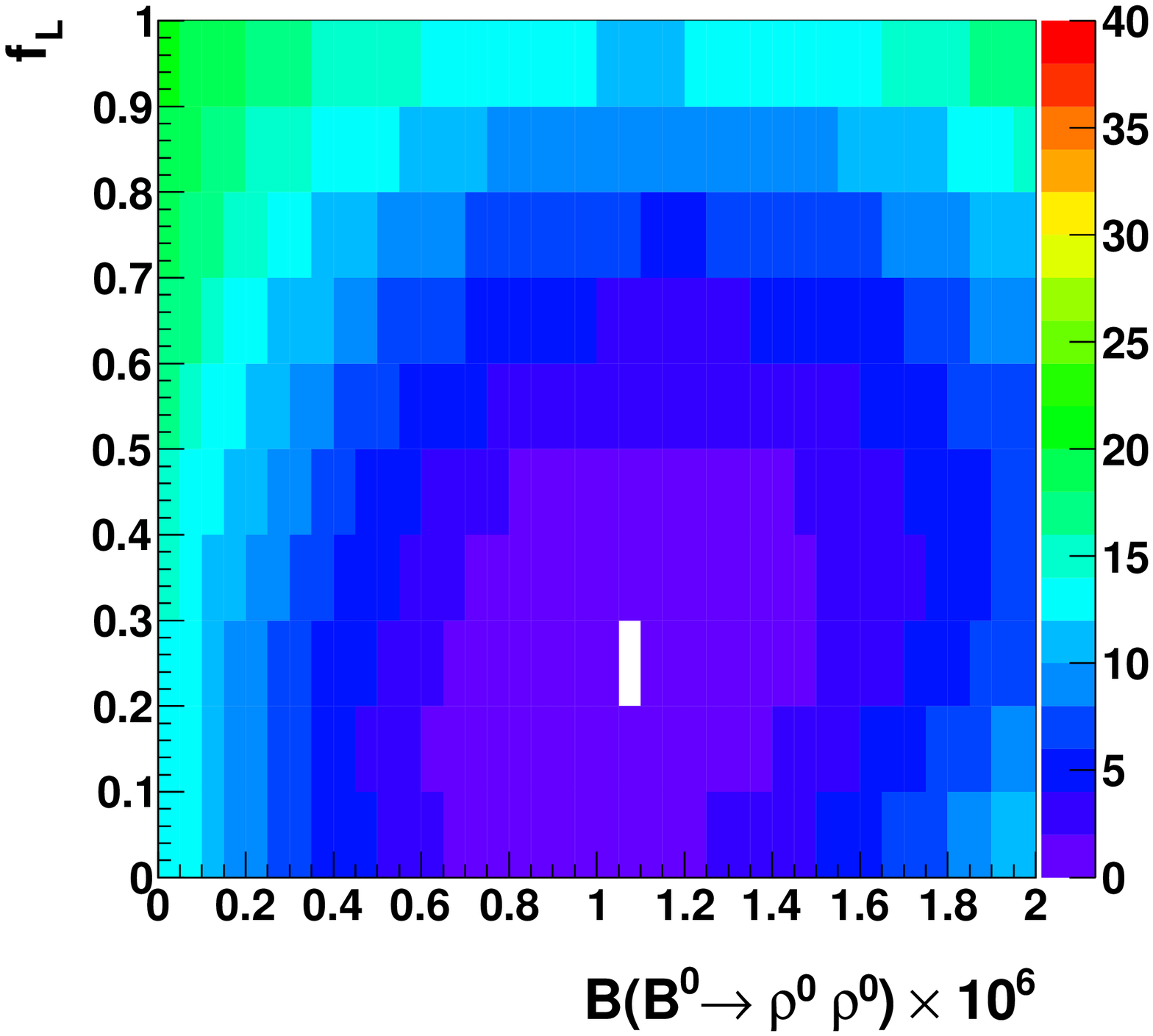}
 \includegraphics[height=!,width=0.32\columnwidth]{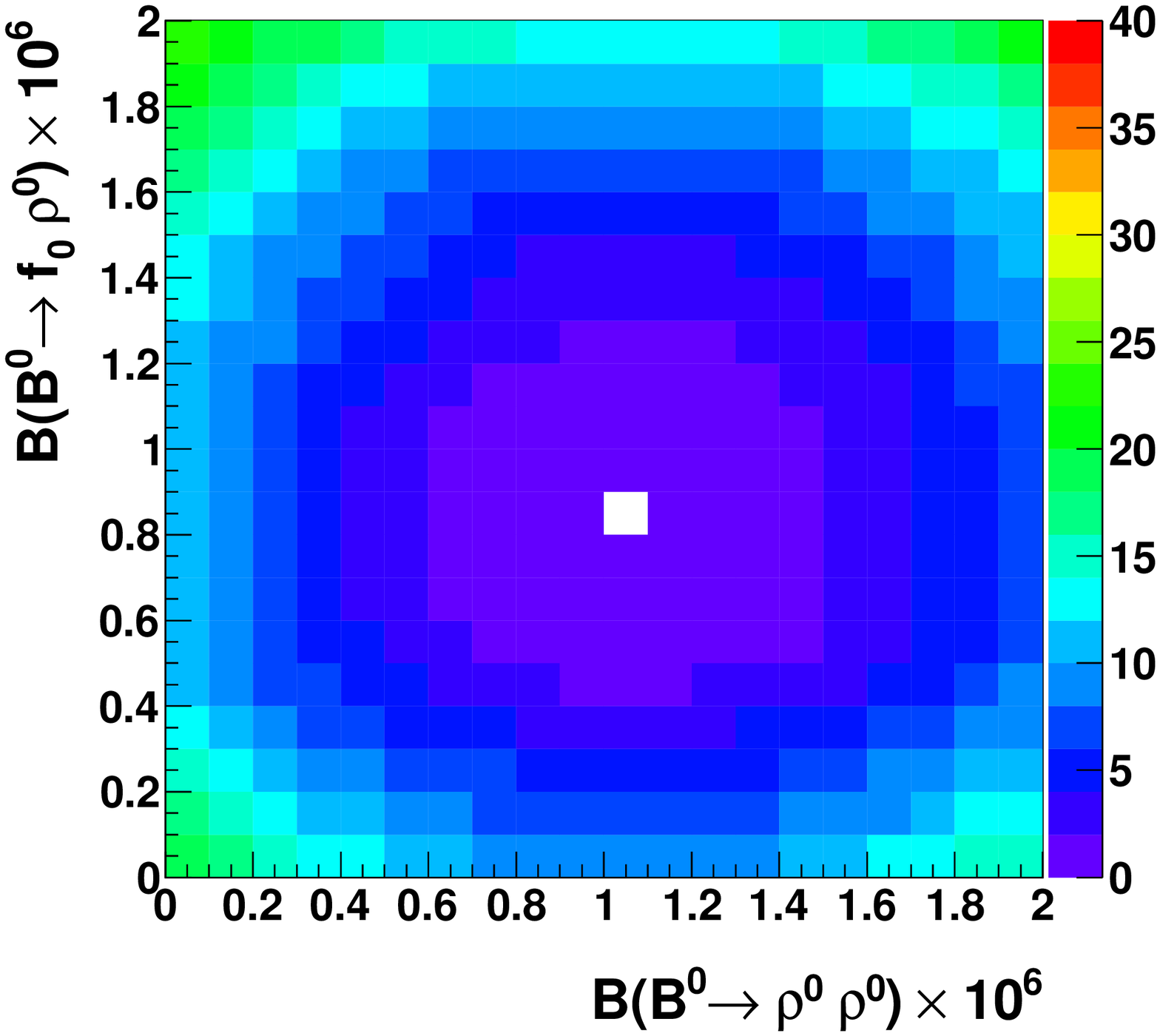}                                                  
\caption{(color online) Top: Likelihood scans of ${\cal B}(\Brr)$, $f_L$ and ${\cal B}(B^0\rightarrow f_0\rz)$ convolved with the corresponding systematic uncertainty.
Bottom: 2D likelihood scans of ${\cal B}(\Brr)$ versus $f_L$ (left and middle) and versus \Bfr\ (right). The statistical likelihood scan of ${\cal B}(\Brr)$ versus $f_L$ (left) demonstrates the expected insensitivity of the likelihood to $f_L$ for ${\cal B}(\Brr)=0$. The middle and right scans are convoluted with the corresponding systematic uncertainties. The minimum (white spot) presents the fit result obtained from this measurement.}  
 \label{pic_LH1}                                                                                                                                              
 \end{figure}

We evaluate the statistical significance ${\cal S}_{0}$ of the result by taking the ratio of the likelihood of the nominal fit (${\cal L}_{\rm max}$) and of the fit with the signal yield fixed to zero (${\cal L}_{0}$);
\begin{eqnarray}
 {\cal S}_{0} = \sqrt{-2\ln\biggl(\frac{{\cal L}_{0}}{{\cal L}_{\rm max}}\biggr)}.
\label{e_sign}
\end{eqnarray}
The statistical significances of the \Brr\ and \Bfr\ yields are found to be $4.6\sigma$ and $3.6\sigma$, respectively. In addition, we perform a likelihood scan for each measured four-pion final state as well as $f_L$, where no exceptional behavior is found. Likelihood scans of ${\cal B}(\Brr)$, $f_L$ and ${\cal B}(\Bfr)$, where the likelihoods are convolved with the a Gaussian whose width is set to the corresponding systematic uncertainty, are shown in Fig.~\ref{pic_LH1}; the one-dimensional scans are used to obtain a total significance of $3.4\sigma$ and $3.1\sigma$ for ${\cal B}(\Brr)$ and ${\cal B}(\Bfr)$, respectively.  

\clearpage 

\section{Validity Checks}
\label{Validity Checks and Significance} 

We have validated our fitting procedure using a full GEANT MC simulation. Within the statistical error, the fitter reliably recovers the input branching fractions for \Brr, \Brpp, \Bpppp, \Bff\ and \Bfpp. For \Bfr, the fitter exhibits a small bias; this is described in Sec.~\ref{Systematic Uncertainties}. 
In order to determine the data-to-simulation correction factors (see Sec.~\ref{Event Model}), we have performed a more limited fit to a control sample of $B^0\to D^-\pi^+, D^- \to K^+ \pim \pim$ decays, which are topologically similar to \Brr. This fit uses only observables \De\ and \Fsb\ for each $r$-bin. The result is ${\cal B}(B^{0}\to D^{-}\pip) = (2.626 \pm 0.015\;(\rm stat)) \times 10^{-3}$, which is in good agreement with the world average value of $(2.68 \pm 0.13)\times 10^{-3}$~\cite{PDG} as the systematic uncertainty is typically of the order of $10\%$.

\begin{figure}
  \centering
  \includegraphics[height=150pt,width=!]{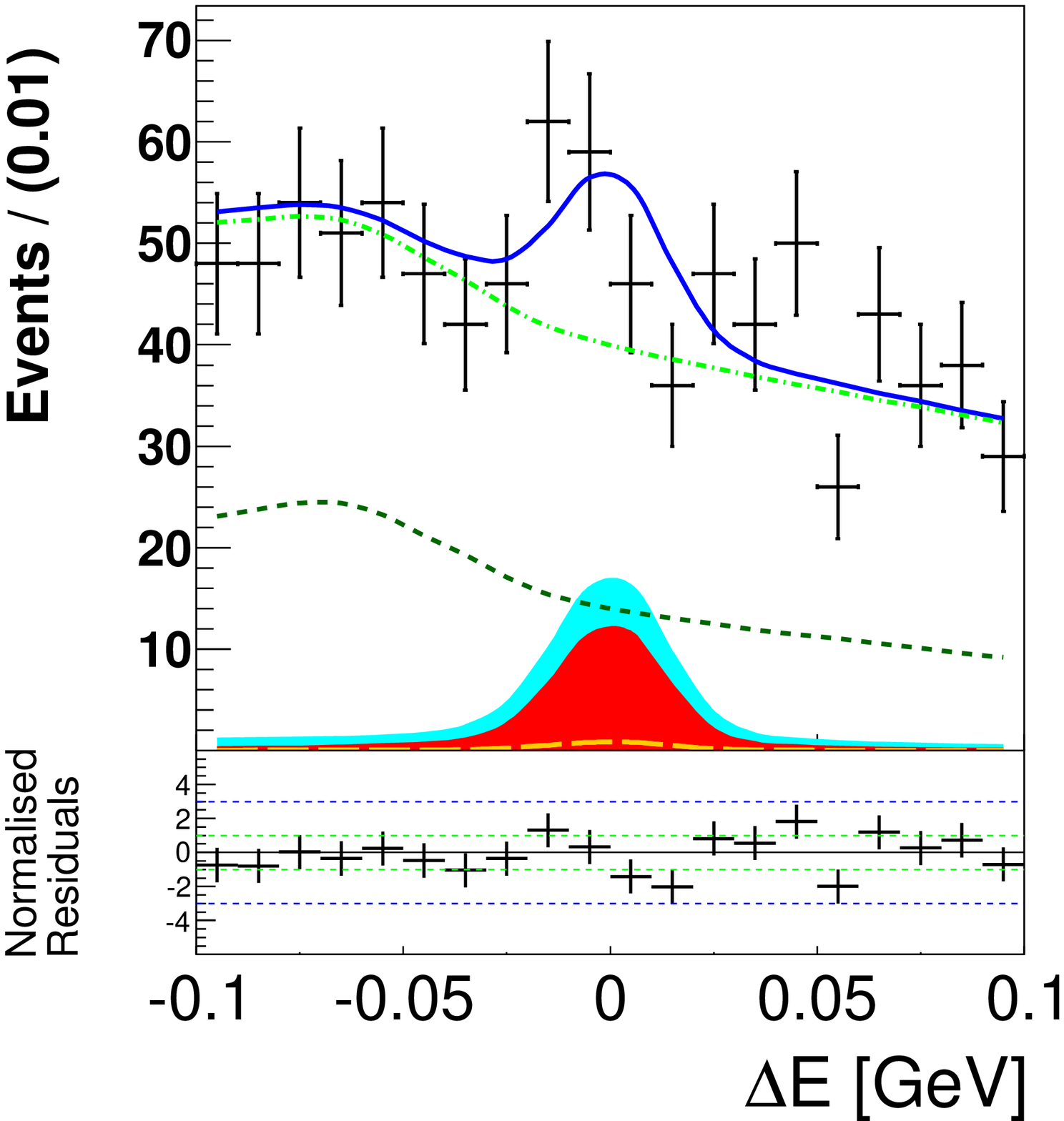}
    \includegraphics[height=150pt,width=!]{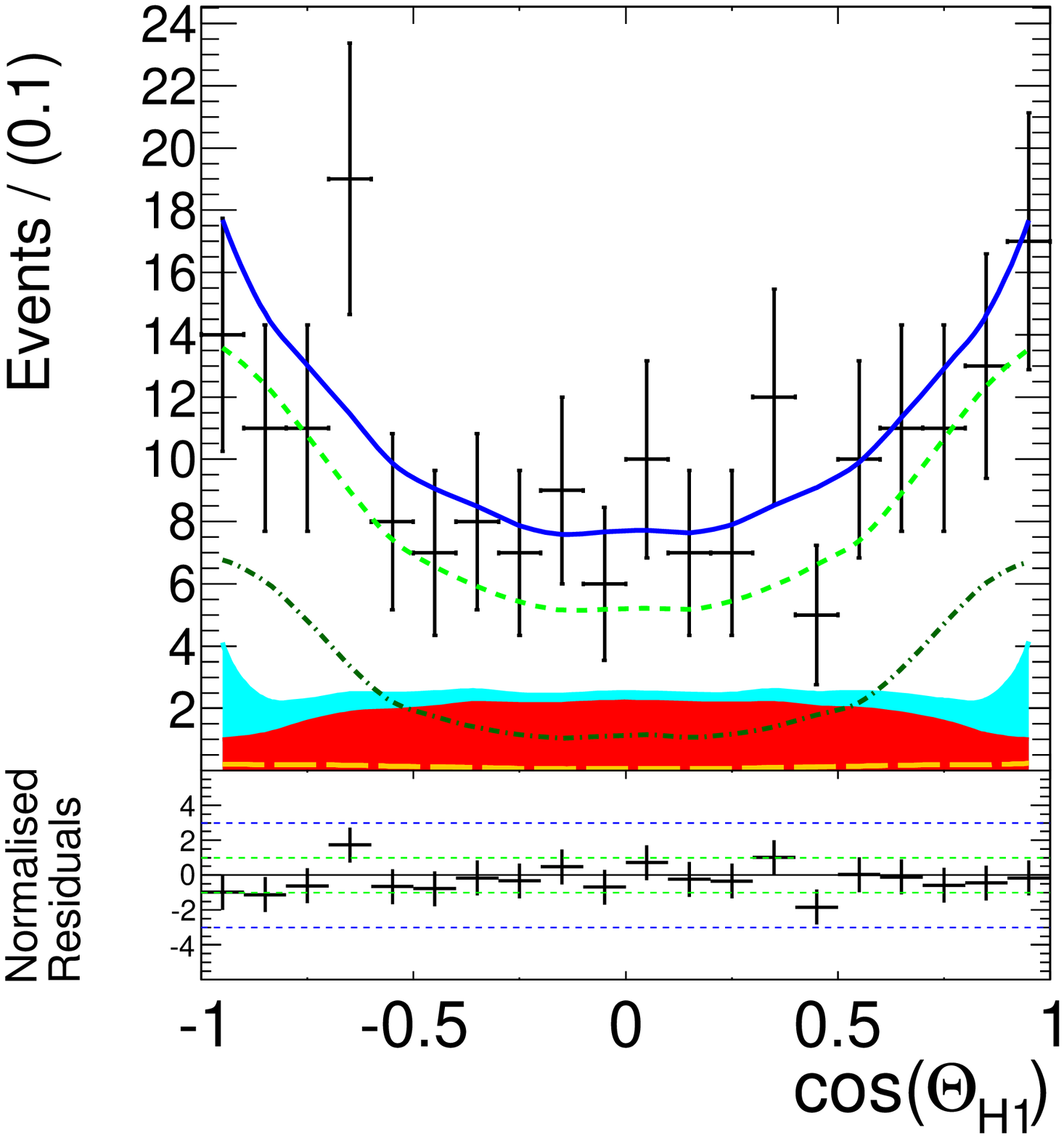}
 \includegraphics[height=150pt,width=!]{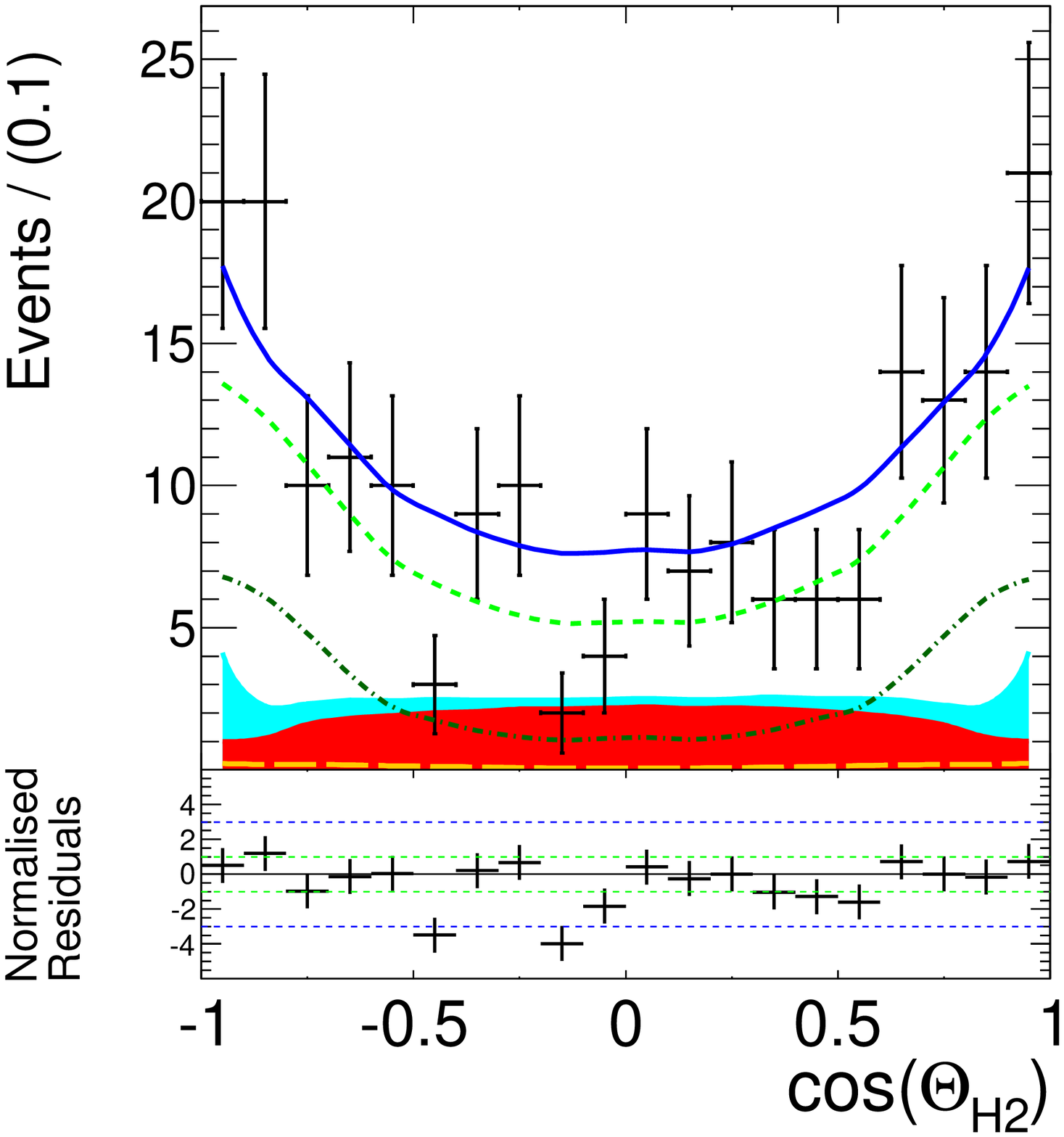}
\put(-350,130){(a)}
  \put(-195,130){(b)}
 \put(-35,130){(c)}

 \includegraphics[height=150pt,width=!]{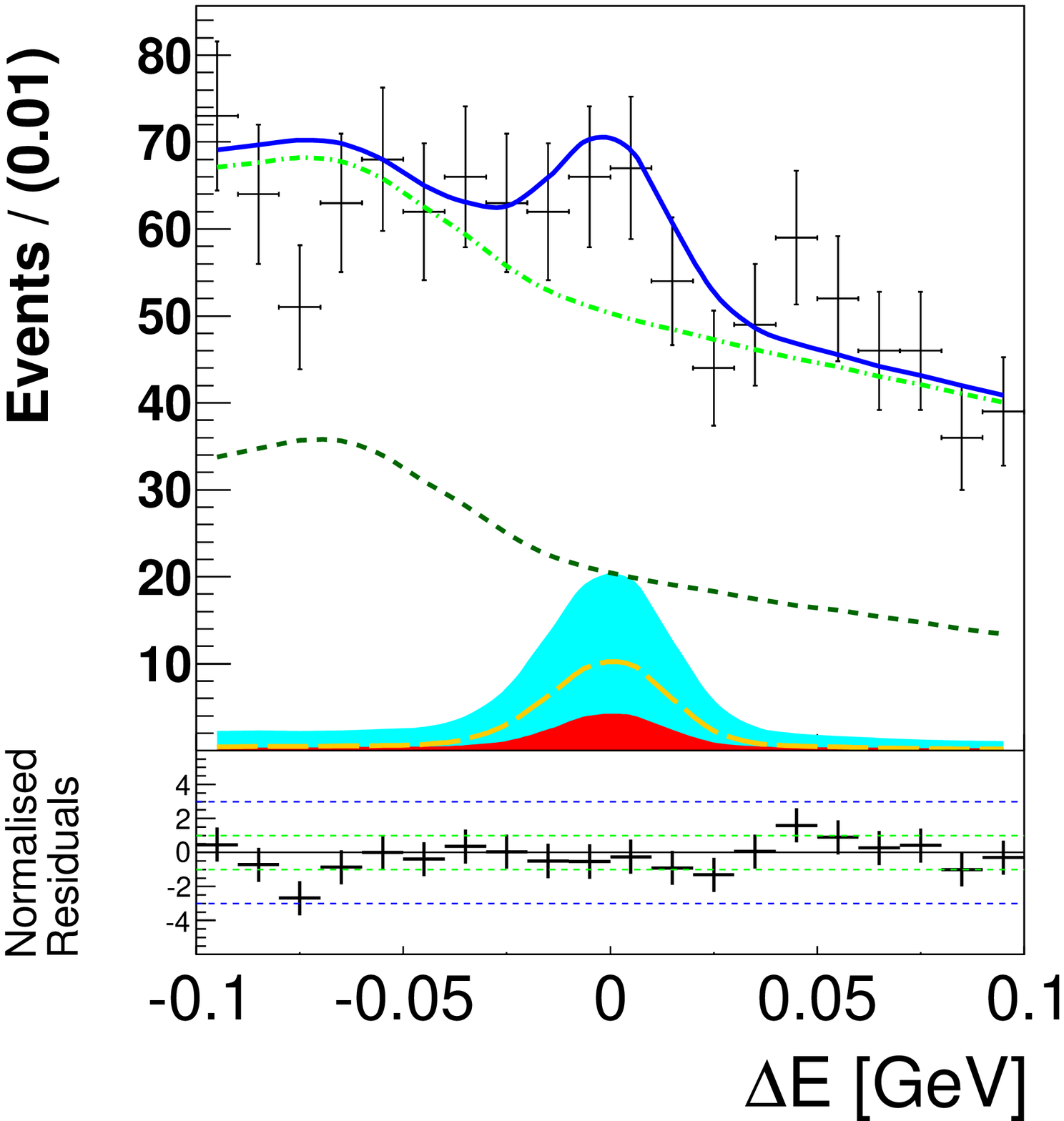}
  \includegraphics[height=150pt,width=!]{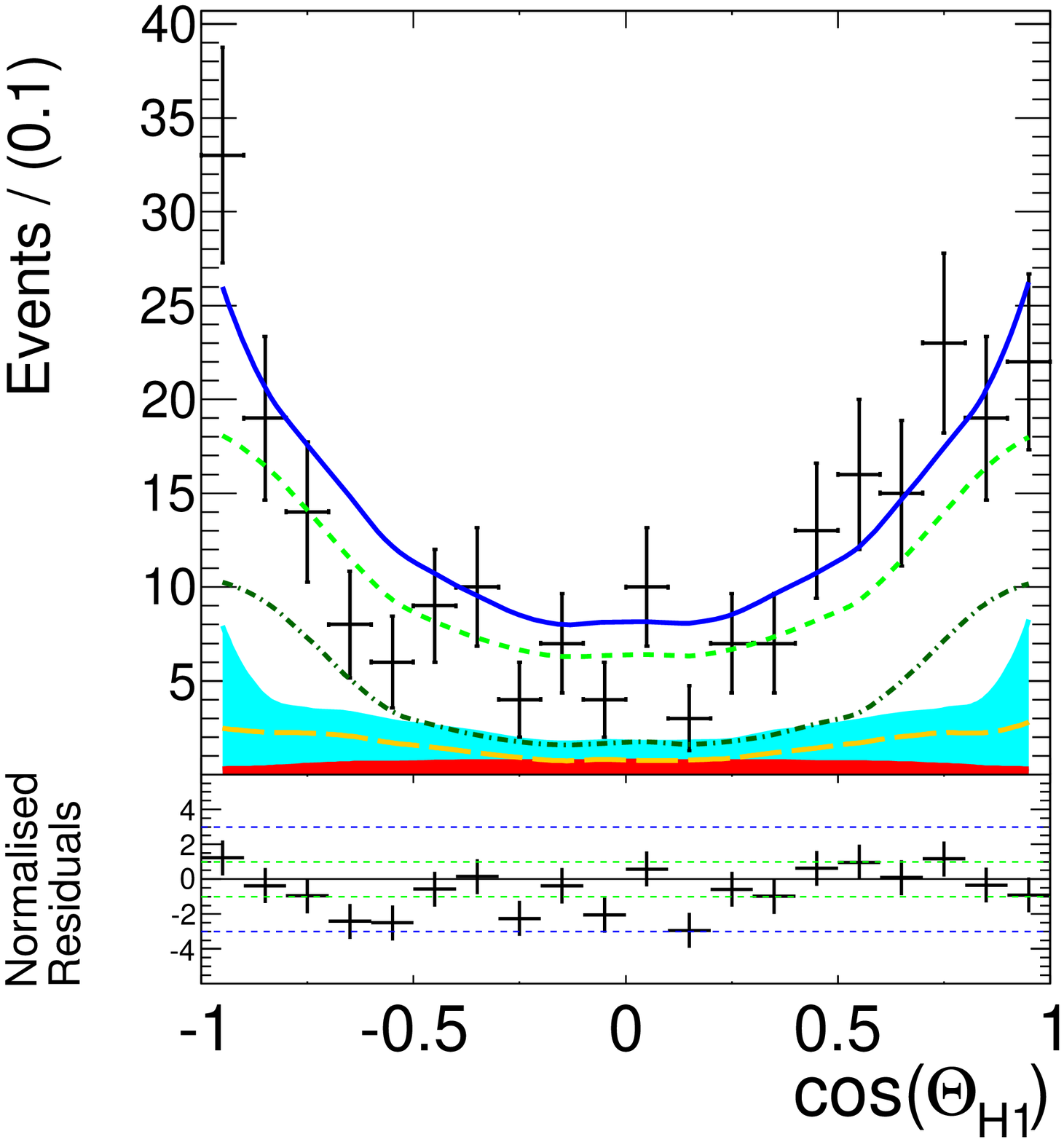}
 \includegraphics[height=150pt,width=!]{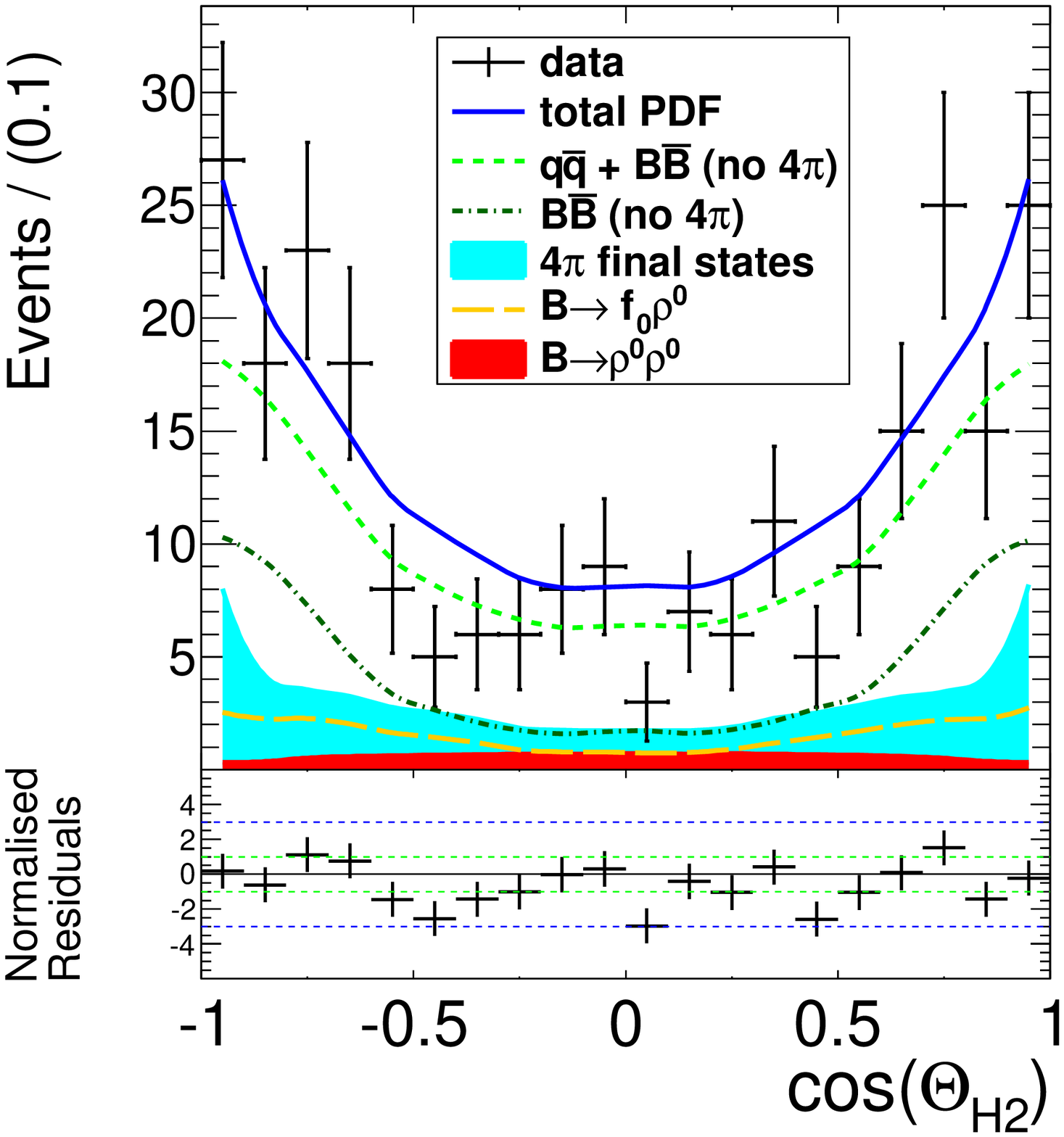}
 \put(-350,130){(d)}
  \put(-195,130){(e)}
 \put(-35,130){(f)}

  \caption{(color online) Projections of the fit to the data into a \rz\rz\ window $0.60\; {\rm GeV/c^{2}} < \Mpp < 0.88\;{\rm GeV/c^{2}}$(upper row) and a $f_0\rz$ window  $0.6\; {\rm GeV/c^{2}} < M_{k} < 0.88\;{\rm GeV/c^{2}}$ and $0.88\; {\rm GeV/c^{2}} < M_{l} < 1.08\;{\rm GeV/c^{2}}$(lower row). The points represent the data and the solid curves represent the fit result. The shaded (red) areas show the \Brr\ and the (orange) long dashed curves the \Bfr\ contribution. The bright shaded (cyan) areas show all four-pion final states, the (dark green) dashed curves show the non-peaking $B\bar{B}$ contribution and the (bright green) dash-dotted curves show the total non-peaking background. Panels (a) and (d) show the projection into \De\ for $\Fsb > 0.5$ and (b,c) and (e,f) show into \cH\ for $\Fsb > 0.5$ and $|\De| < 0.02 \; {\rm GeV}$. The residuals are plotted below each distribution.}
  \label{fig_bf_data2}
\end{figure}

We investigate the stability of the fit result by removing from the fit the components $B^{0}\rightarrow \pip\pim\pip\pim$, $f_0\pip\pim$, and $f_0f_0$, whose yields in the nominal fit are negative, consistent with zero. Separately, we furthermore remove $B^{0}\rightarrow \rho^{0}\pip\pim$. The fit result remains stable in both scenarios.
Since the branching fraction of $B^{0}\rightarrow f_0\rz$ is larger than indicated by the previous measurement, we investigate the impact on $\Brr$ when setting ${\cal B}(B^{0}\rightarrow f_0\rz) = 0$ and vice versa. We obtain consistent results,: $\Brr|_{f_0\rz = 0}$ = $ (1.01 \pm 0.31\;(\rm stat))\times 10^{-6}$ with $f_L|_{f_0\rz = 0} = 0.26 \pm 0.21\;(\rm stat) $ and $\Bfr|_{\rz\rz = 0} = (0.73 \pm 0.21\;(\rm stat))\times 10^{-6}$. In order to visualize each mode separately, the nominal fit projections into the $\rz\rz$ and $\rz f_0$ windows are shown in Fig.~\ref{fig_bf_data2}, where in both $\De$ distributions a signal peak is visible, while the excess in the center region of the helicity angle distributions in the $\rz\rz$ window indicates the preferred transverse polarization. 

Next, we refit the data while fixing $f_L$ to either $0$ or $1$. We obtain ${\cal B}(\Brr)_{f_L = 0} = (0.79 \pm 0.22\;(\rm stat))\times 10^{-6}$ and ${\cal B}(\Brr)_{f_L = 1} =(0.22 \pm 0.24\;(\rm stat))\times 10^{-6}$, respectively. The events from the polarization fixed to zero prefer an assignment to the background over a modified polarization. In addition to the two invariant dipion masses, the two helicity angles provide additional separation power (especially for the four-pion final state), which in some cases can be larger than the one from the masses, as many backgrounds have a $\rho^{0}$ contribution. We demonstrate their importance by removing the helicity angles from the fit where we obtain  ${\cal B}(\Brr) = 0.3^{+0.42}_{-0.31}\;(\rm stat)\times 10^{-6}$ and ${\cal B}(\Bfr) = (0.29 \pm 0.28 \;(\rm stat))\times 10^{-6}$, while we see an excess ($<2\sigma$) in the modes \Bpppp\ and \Brpp. This is consistent with the previous analysis from Belle~\cite{r0r0_Belle}.

\section{Systematic Uncertainties}
\label{Systematic Uncertainties} 
Systematic errors from various sources are considered and estimated with independent studies and cross-checks. These are summarized in Table~\ref{tab_bf_syst} for \Brr\ and Table~\ref{t_syst} for the remaining four-pion modes. This includes the uncertainty on the number of produced \BBbar\ events in the data sample. Contributions to the uncertainty in the selection efficiency due to particle identification and tracking are determined by using independent control samples.

The uncertainty in the \rz\ shape is determined by varying the fixed mass and width within its world average uncertainty~\cite{PDG}.
We account for a difference in the fraction of mis-reconstructed events between data and MC by varying this parameter by $\pm50\%$ of its value and repeating the fit. Since we did not treat the correctly reconstructed events separately for the remaining four pion final states, we assign the same number as for $\Brr$. 

Variations in the parametric model shape due to limited statistics are accounted for by varying each parameter within its error. Uncertainties in the non-parametric shapes are obtained either by varying the contents of the histogram bins within $\pm 1\sigma$ (referred to as bin-wise in Table~\ref{tab_HMvariation_syst}) or by modifying the shapes as follows; in order to enhance a longitudinally (transversely) polarized signal-like shape we vary the content of each two-dimensional helicity PDF from $-$($+$) $1\sigma$ in the center and $+$($-$) $1\sigma$ in the corners and extrapolate the variation in between. For each background component, we generate two sets of pseudo-experiments according to our fit result; one for each modification of the helicity PDFs (longitudinal- or transverse-like). We neglect the non-significant modes in this study, except for \Brpp, whose yield is free; the variation of the \Brpp\ helicity shape is described below.  Half of the difference between the fitted parameter of each set is taken as the corresponding uncertainty from this study (referred to as bending in Table~\ref{tab_HMvariation_syst}). Since the helicity description of continuum is fixed from off-resonance data, we also use sideband data ($M_{\rm bc}<5.25\; \rm GeV/c^{2}$ and $\De > 0.025\; \rm GeV$) in order to correct the continuum helicity PDF in each bin, apply the correction to the fit to the data and take the difference to the nominal fit result as the corresponding uncertainty. Similarly, we obtain an alternative, overall description of the helicity and mass distributions of generic $B$ decays from another sideband region ($M_{\rm bc}<5.25\; \rm GeV/c^{2}$ and $|\De| < 0.2 \;\rm GeV$), which then substitutes the one obtained from MC simulation in the fit to data. To avoid double-counting of the same effect, the largest uncertainty of all three studies, see Table~\ref{tab_HMvariation_syst}, is taken. We consider the diversity of all three studies as large enough to also account for a possible difference between data and MC.

The systematic uncertainty due to fixing the peaking background yields of \aonepi\ and \bonepi\ are estimated by varying the branching fraction by its world average error and repeating the fit. Since only an upper limit is known for \atwopi, we vary its yield from zero to two times its fixed value.  
The fit bias was determined from full simulation by searching for a difference between the generated and fitted physics parameters. Because of imperfection in the modeling of all the correlations, we find a non-negligible bias of $+16\%$ for the mode $B^{0}\to f_0\rz$. We subtract $16\%$ from the fit result and assign a $2.2\%$ uncertainty, determined from a variation of the generated  $B^{0}\to f_0\rz$ yield within $\pm 1\sigma$. All other biases are found to be small compared to the statistical uncertainty and are therefore treated fully as systematic uncertainties.
Furthermore, we performed an ensemble test where we replaced the \Brpp\ helicity PDF with one where the \rz\ is either longitudinally or transversely polarized to generate MC sets according to the fit result. Modes in agreement with zero events except \Brpp were not generated but left free in the fits. The maximal deviation from the nominal model is taken as the uncertainty related to the assumption of the \Brpp\ helicity dependency.

Finally, the uncertainty from neglecting interference between the four-pion final states is estimated by constructing a 4-body amplitude and generating samples of two four-pion final states, including detector effects. For each set of modes, we first calibrate the relative amplitude strength between  the two considered modes in order to obtain a yield ratio as found in the data. For the calibration, we set the relative phase to $90^{\circ}$. Then, we generate sets with the relative phase between the two modes of interest varying from $0^{\circ}$ to $180^{\circ}$ in steps of $10^{\circ}$. Each set is fitted with an incoherent model and the RMS of the variation of the fit results with respect to the one obtained from the calibration set is taken to be the systematic uncertainty. We consider the modes $B^{0}\to \rz\rz, a_1\pi, f_0\rz$ and $\rz\pi\pi$ (where, for $\rz\rz$, we set $f_L = 0.21$). We find that interference with $a_1\pi$ gives the largest uncertainty in all cases.

Due to the combination of a broad (non-resonant pion pair) part together with a \rz\ contribution, the mode \Brpp\ can absorb changes in model more easily; therefore this mode has a relatively large statistical and systematical uncertainty.
\begin{table}
  \centering
  \caption{Systematic uncertainties of the branching fraction of \Brr\ and $f_L$.}
  \begin{tabular}
    {@{\hspace{0.5cm}}c@{\hspace{0.25cm}}  @{\hspace{0.25cm}}c@{\hspace{0.25cm}}  @{\hspace{0.25cm}}c@{\hspace{0.5cm}}}
    \hline \hline
    Category & $\delta{\cal B}(\Brr)$ $(\%)$ & $\delta f_L$\\
    \hline
    $\delta N(\BBbar)$ & 1.4 & n.a.\\
    Tracking & 1.4 & n.a.\\
    Particle identification & 2.5 & n.a.\\
    Mis-reconstruction fraction & 2.4 & 0.03\\
    Resonance shape & 0.2 &  $<0.001$\\
    Model shape & 5.1 & 0.11\\
Histogram shape & 8.5 & 0.08\\
    ${\cal B}(B^{0}\to a_1\pi)$ & 0.4 &0.03 \\
    ${\cal B}(B^{0}\to b_1\pi)$ & $<$0.1 &$<0.001$\\
    ${\cal B}(B^{0}\to a_2\pi)$ &$<$0.1 & $<0.001$\\
    Fit bias & 1.9 & 0.03\\
 $\rz\pi\pi$ helicity & 6.3 & 0.05 \\
   Interference & 8.4 & 0.03 \\
\hline
     Total & 15.1 & 0.15\\
    \hline \hline
  \end{tabular}
  \label{tab_bf_syst}
\end{table}

\begin{table}[h]                                                                                                                                                               
\centering    
   \caption{Systematic uncertainties ($\%$) for other modes with four-pion final state.}        
 \begin{tabular}{@{\hspace{0.5cm}}c@{\hspace{0.5cm}}  @{\hspace{0.5cm}}c@{\hspace{0.5cm}} @{\hspace{0.5cm}}c@{\hspace{0.5cm}} @{\hspace{0.5cm}}c@{\hspace{0.5cm}} @{\hspace{0.5cm}}c@{\hspace{0.5cm}} @{\hspace{0.5cm}}c@{\hspace{0.5cm}}}
\hline \hline 
\small Category & 4$\pi$ &$\rz\pip\pim$ & $f_0\pip\pim$ & $f_0f_0$ & $f_0\rz$\\
\hline \hline 
$\delta N(B\bar{B})$ & 1.4 & 1.4 & 1.4 & 1.4 & 1.4 \\
Tracking & 1.4 & 1.4 & 1.4 & 1.4 & 1.4\\
 Particle identification & 2.5 & 2.5 & 2.5 & 2.5 & 2.5\\
Mis-reconstruction fraction & 2.4 & 2.4 & 2.4 & 2.4 & 2.4\\
Resonance shape & n.a. &n.a. &n.a. &n.a. & $<1$\\
Model shape & 28.5 & 218.8 & 13.5 & 13.8 & 7.1\\
Histogram shape & 38.1 & 127.3 & 54.5 & 46.1 & 5.9\\

 ${\cal B}(B^{0}\to a_1\pi)$ & 10.3 & 129.5 & 3.1 & 4.7 & 3.4\\
 ${\cal B}(B^{0}\to b_1\pi)$ & $<$1 & 1.6 & $<$1 & $<$1  &  $<$1\\
 ${\cal B}(B^{0}\to a_2\pi)$ & $<$1 & 2.7 & $<$1 &  $<$1 &  $<$1\\ 
Fit bias & 18.6 & 10.3 & 7.4  & 100.1 & 2.2 \\
$\rz\pi\pi$ helicity & 26.8  & 23.3 & 17.6 & 14.1 & 4.5 \\
Interference & n.a. & 93.2 & n.a. & n.a. & 6.8\\
 \hline \hline
Total & 58.7 & 300.3 &59.5  &112.1  &13.7 \\
\hline \hline  
 \end{tabular}                                          

\label{t_syst}                                                                                                                                                           
\end{table}

\begin{table}
  \centering
  \caption{Systematic uncertainties arising from the three alternative studies of the variation of the non-parametric PDFs as described in the text. The largest value is taken for each observable. The uncertainties obtained from the sideband studies are listed separately in the last two columns. For the observable ${\cal B}(\Brr)$ we combine the value obtained from correcting the continuum helicity PDF with sideband data with the remaining values obtained from the bending studies ($6.2\%$), resulting in a total uncertainty of $8.5\%$. The $B\bar{B}$ components give the dominant contribution to the uncertainty of $f_L$ in all cases.}
  \begin{tabular}
    {@{\hspace{0.5cm}}c@{\hspace{0.25cm}} |  @{\hspace{0.25cm}}c@{\hspace{0.25cm}} @{\hspace{0.25cm}}c@{\hspace{0.25cm}} @{\hspace{0.25cm}}c@{\hspace{0.25cm}}  @{\hspace{0.25cm}}c@{\hspace{0.5cm}}}
 \hline \hline
 Variation & Bin-wise & Bending &  $q\bar{q}$ sideband &  $B\bar{B}$ sideband \\
\hline
$\delta{\cal B}(\Brr)$ $(\%)$ & 4.1 & 7.0 & 5.8 & 0.4\\
 $\delta f_L$ & 0.03 & 0.08 & 0.05 & 0.07\\
$\delta{\cal B}(\Bfr)$ $(\%)$ & 5.9 & 3.4 & 2.6 & 3.8\\
 \hline \hline
  \end{tabular}
  \label{tab_HMvariation_syst}
\end{table}

\section{$\phi_{2}$ constraint}
\label{phi2 constraint}
\begin{figure}[h]                                                                                                                                                               
\centering                                                                                                                                                                     
\includegraphics[height=!,width=0.6\columnwidth]{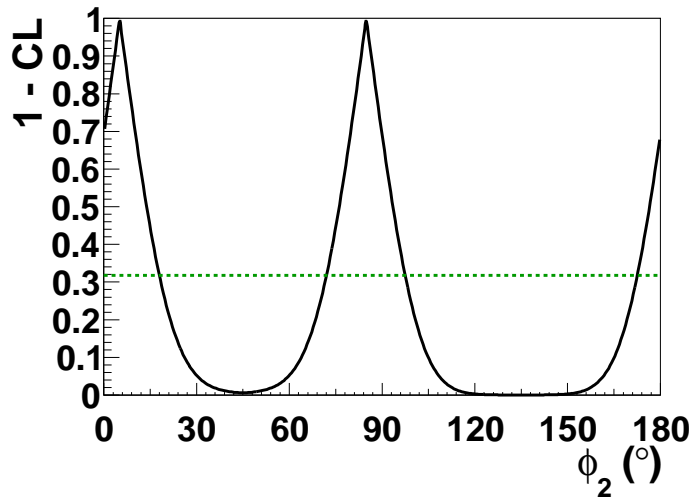}
\caption{$1-$CL versus $\phi_2$ obtained from an isospin analysis from $B\to\rho\rho$ decays. The horizontal line shows the $68\%$ CL.}
\label{p_phi2}                                                                                                                                                   
\end{figure}

We use the branching fraction and the longitudinally polarized fraction of \Brr\ decays from our result; $f_L\times{\cal B}(\Brr) = (0.21 \pm 0.34)\times10^{-6}$ to obtain a new constraint on the CKM angle $\phi_{2}$ through an isospin analysis~\cite{theory_isospin} in the $B\to\rho\rho$ system. Because of Bose-Einstein statistics, the two $\rho$s can only carry a total isospin of $I=0$ or $I=2$ while the strong loop contributions can only result in $I=0$; the gluon does not carry isospin. Neglecting electroweak contributions or isospin breaking effects, the complex $B\to\rho\rho$ amplitudes can be related via
\begin{equation}
\frac{1}{\sqrt{2}}A^{+-} + A^{00} = A^{+0}, \;\;\;\;\;\;\;\;\frac{1}{\sqrt{2}}\bar{A}^{+-} + \bar{A}^{00} = \bar{A}^{-0},
\end{equation}
where the amplitudes with $\bar{b}\to \bar{u}$ ($b \to u$) transitions are denoted as $A^{ij}$ ($\bar{A}^{ij}$) and the superscript identifies the charges of the $\rho$ mesons. These relations can be visualized as two triangles in the complex plane. Isospin arguments show that the charged $B$ decay $B^{\pm}\to\rho^{\pm}\rho^0$ arises only at tree level. Consequently, the two isospin triangles share the same base: $A^{+0} = \bar{A}^{-0}$. The difference between the two isospin triangles corresponds to the shift $\Delta \phi_2$ due to additional contributions. This method has a 8-fold ambiguity in the determination of $\phi_2$ that arises from the four possible orientations of the two triangles and from measuring $\sin(\phi_2^{\rm eff})$. The amplitudes are constructed from branching fractions and direct $CP$ asymmetries ${\cal A}_{CP}$ and then used to obtain the possible pollution in the mixing induced $CP$ asymmetry ${\cal S}_{CP} = \sqrt{1-A_{CP}^{2}} \sin(2\phi_{2}^{\rm eff})$, obtained from $B^0\to\rho^+\rho^-$ decays. For the remaining sides of the triangles, we use Belle results: the longitudinally polarized fraction of ${\cal B}(B^{0}\to \rho^{+}\rho^{-}) = (22.8 \pm 4.6) \times 10^{-6}$ with $f_L^{+-} = 0.94\pm 0.05$, ${\cal A}^{+-}_{CP} = 0.16 \pm  0.22$, ${\cal S}^{+-}_{CP} = 0.19 \pm 0.31$~\cite{rhorho_Belle, Brprm_Belle} and the longitudinally polarized fraction of ${\cal B}(B^{\pm}\to \rho^{\pm}\rho^{0}) = (31.7 \pm 8.8)\times 10^{-5}$ with $f_L^{+0} = 0.95\pm0.11$~\cite{rpr0_Belle}. Figure~\ref{p_phi2} shows the two solutions for $\phi_2$ from a probability scan. The solution that is consistent with the SM is $\phi_{2} = (84.9 \pm 13.5)^{\circ}$. The size of the penguin contributions is small: $\Delta \phi_2 = (0.0 \pm 10.4)^{\circ}$. Because of the very small $\Brr$ branching fraction relative to the other two $B\to\rho\rho$ decays, the four solutions from the isospin analysis degenerate into the two apparent solutions. This makes this isospin analysis less ambiguous compared with the $B\to\pi\pi$ system, where the decay into two neutral pions is significant stronger~\cite{pi0pi0_Belle, pi0pi0_Babar, pipi_Belle, pipi_BABAR}, resulting in eight solutions for $\phi_2$. 

\section{Result Discussion}
\label{Result Discussion}

Including the helicity angles in the fit is the main difference with the previous Belle analysis, where no such information was used. Besides allowing us to measure the polarization, the helicity angles provide further separation power, especially between the four-pion final states, which otherwise can only be separated by the dipion masses. Their importance is reflected by the result of the fit to data without using the helicity information (see Section~\ref{Validity Checks and Significance}). The stability of the fit result within the model shape studies (see Section~\ref{Systematic Uncertainties}) demonstrates the use of an appropriate helicity description. Further differences include an improved tagging algorithm for the SVD2 data sample, the usage of \Mbc\ as a $B$ meson selection criteria instead of the vertex fit quality (so that this measurement could be superseded by a time-dependend measurement; consequently, \Mbc\ could not be used as a fit variable here, as mentioned in Sec.~\ref{Event Selection}) and the replacement of a event-shape topology-dependent selection criteria with the inclusion of $\Fsb$ into the fit. The latter increases the amount of continuum background significantly more than $B\bar{B}$ mesons decay contributions; to compensate, reduced $M_1$, $M_2$ window has been chosen to reduce the overall background while leaving the signal detection quality unchanged, as demonstrated with a MC study. Moreover, an optimization of the selection criteria, but especially the inclusion of $\Fsb$, results in an increase of the reconstruction efficiency by $107\%$, according to MC simulation. For the comparison, we assume the same $f_L$ as obtained from this measurement for the total reconstruction efficiency of the previous analysis, since the efficiency for transversely polarized $\rz$s is higher due to a different momentum spectrum of the daughter pions.

\section{Conclusion}
\label{Conclusion}
We have presented a measurement of the branching fraction of \Brr\ decays and the fraction of longitudinally polarized $\rho$ mesons in this decay, together with other four-pion final states using the final Belle data set of $772\times 10^{6}$ \BBbar\ pairs. 

We find a branching fraction of ${\cal B}(\Brr) = (1.02\pm 0.30\;(\rm stat)  \pm 0.15\;(\rm syst))\times 10^{-6} $ with a significance of $3.4$ standard deviations and a longitudinally polarization fraction $f_L = 0.21^{+0.18}_{-0.22} \;(\rm stat) \pm 0.15 \;(\rm syst)$. Since the longitudinally polarization fraction is found to be small, no measurement of the $CP$ asymmetries is performed. However, we use the result of longitudinally polarized $\rho$ mesons in \Brr\ decays to constrain the CKM angle $\phi_2 =  (84.9 \pm 13.5)^{\circ}$ with an isospin analysis in the $B\to\rho\rho$ system. 

Furthermore, we find $125 \pm 41$ \Bfr\ events, corresponding to ${\cal B}(\Bfr) \times {\cal B}(\Fpp) = (0.78 \pm 0.22 \; (\rm stat) \pm 0.11 \;(\rm syst))\times 10^{-6}$, with a significance of $3.1$ standard deviations. With a significant yield of \Bfr\ decays,
 a measurement of the $CP$ asymmetries could be performed in principle but with a large uncertainty with the current statistics. We find no other significant contribution with the same final state, and set upper limits at 90\% confidence level on the (product) branching fractions, ${\cal B}(\Bpppp) < 11.2 \times 10^{-6} $, ${\cal B}(\Brpp) < 12.0 \times 10^{-6}$, ${\cal B}(\Bfpp) \times {\cal B}(\Fpp) < 3.0 \times 10^{-6}$ and ${\cal B}(\Bff) \times {\cal B}(\Fpp)^{2} < 0.2 \times 10^{-6}$.

The previous Belle analysis set an upper limit on the branching fraction of \Brr\ decays; ${\cal B}(\Brr) <1.0 \times 10^{-6}$ at $90\%$ confidence level (CL), using a sample containing $657 \times 10^{6}$ \BBbar\ pairs and assuming pure longitudinal polarization.~\cite{r0r0_Belle}. The BaBar collaboration has performed a study of \Brr\ decays with $465 \times 10^{6}$ \BBbar\ pairs and found a branching fraction ${\cal B}(\Brr) = (0.92 \pm 0.32\;(\rm stat)\pm0.14 \; (\rm syst))\times 10^{-6}$ and a longitudinal polarization fraction of $f_L = 0.75^{+0.11}_{-0.14}\;(\rm stat)\pm0.04\;(\rm syst)$~\cite{r0r0_BABAR}. Thus the resulting $B^{0}\to\rz\rz$ branching fraction is consistent with the previous Belle analysis and it is also in agreement with the value obtained by the BaBar collaboration. The fraction of longitudinal polarization in \Brr\ decays is somewhat lower than previously measured (differing from the BaBar result by $2.1\sigma$) and the branching fraction of $B^{0}\to f_0\rz$ decays is significantly higher than indicated by previous measurements. For the modes where an upper limit is obtained, we have improved values for \Bpppp\ and \Bfpp\ compared to the current available limits~\cite{PDG}.

This measurement is still statistically limited and more insight on the interesting and complex structure of four-pion final state $B$ decays will be accessible by future experiments~\cite{belle2, LHCb}, e.g, by performing a four-body Dalitz analysis.

\section*{Acknowledgments}

We thank the KEKB group for the excellent operation of the
accelerator; the KEK cryogenics group for the efficient
operation of the solenoid; and the KEK computer group,
the National Institute of Informatics, and the 
PNNL/EMSL computing group for valuable computing
and SINET4 network support.  We acknowledge support from
the Ministry of Education, Culture, Sports, Science, and
Technology (MEXT) of Japan, the Japan Society for the 
Promotion of Science (JSPS), and the Tau-Lepton Physics 
Research Center of Nagoya University; 
the Australian Research Council and the Australian 
Department of Industry, Innovation, Science and Research;
Austrian Science Fund under Grant No. P 22742-N16;
the National Natural Science Foundation of China under
contract No.~10575109, 10775142, 10875115 and 10825524; 
the Ministry of Education, Youth and Sports of the Czech 
Republic under contract No.~MSM0021620859;
the Carl Zeiss Foundation, the Deutsche Forschungsgemeinschaft
and the VolkswagenStiftung;
the Department of Science and Technology of India; 
the Istituto Nazionale di Fisica Nucleare of Italy; 
The BK21 and WCU program of the Ministry Education Science and
Technology, National Research Foundation of Korea Grant No.\ 
2010-0021174, 2011-0029457, 2012-0008143, 2012R1A1A2008330,
BRL program under NRF Grant No. KRF-2011-0020333,
and GSDC of the Korea Institute of Science and Technology Information;
the Polish Ministry of Science and Higher Education and 
the National Science Center;
the Ministry of Education and Science of the Russian
Federation and the Russian Federal Agency for Atomic Energy;
the Slovenian Research Agency;
the Basque Foundation for Science (IKERBASQUE) and the UPV/EHU under 
program UFI 11/55;
the Swiss National Science Foundation; the National Science Council
and the Ministry of Education of Taiwan; and the U.S.\
Department of Energy and the National Science Foundation.
This work is supported by a Grant-in-Aid from MEXT for 
Science Research in a Priority Area (``New Development of 
Flavor Physics''), and from JSPS for Creative Scientific 
Research (``Evolution of Tau-lepton Physics'').

\end{document}